\documentclass[aip,reprint,amsmath,amssymb]{revtex4-2}

\usepackage[utf8]{inputenc}
\usepackage[T1]{fontenc}
\usepackage{graphicx}
\usepackage{tikz}
\usetikzlibrary{arrows.meta, positioning, fit, calc}
\graphicspath{{figs/}}   
\usepackage{booktabs}          
\usepackage{siunitx}           
\usepackage{xcolor}
\usepackage[colorlinks=true,linkcolor=blue,citecolor=blue]{hyperref}

\newcommand{\Rey}{\mathrm{Re}}
\newcommand{\St}{\mathrm{St}}
\newcommand{\zvec}{\mathbf{z}}
\newcommand{\evec}{\mathbf{e}}
\newcommand{\uvec}{\mathbf{u}}

\begin{document}

\title{Autoregressive rollout error in latent-space reduced-order models
       of bluff-body wakes is accumulated phase drift}

\author{Suvam Samanta}
\affiliation{{Department of Physics}, National Institute of Technology Agartala,
             Jirania, Tripura 799046, India}

\author{Sachidananda Behera}
\affiliation{Department of Mechanical and Aerospace Engineering,
             Indian Institute of Technology Hyderabad,
             Kandi, Sangareddy 502284, Telangana, India}

\date{\today}

\begin{abstract}
Reduced-order models that advance a low-dimensional latent state autoregressively
are limited by rollout error that compounds over long horizons and is usually
treated as unstructured. We show that for bluff-body wakes it is highly structured,
and that its structure reveals what these models learn. The rollout-error spectrum
of a convolutional-autoencoder--LSTM model is sharply peaked at the vortex-shedding
frequency --- for a circular cylinder over $\Rey = 300$--$800$ and a square cylinder
at $\Rey = 100$, following each flow's own frequency and the error is
$97$--$98\%$ \emph{phase} error (down to $95\%$ across a seed ensemble): the network
reproduces the limit-cycle amplitude to within $0.15\%$ but traverses the cycle at
slightly the wrong rate. It learns the geometry of the attractor almost exactly and
its temporal parametrisation slightly wrong, and this timing error --- a
total of only a few thousandths of a cycle accumulated over the whole rollout produces the entire long-horizon error while one-step validation error and reconstructed fields appear
essentially perfect.

This diagnosis is the central result; the correction follows from it. Because the
phase drifts linearly it is removed by one parameter per latent coordinate fitted
on a short calibration window, with a diagnostic --- the signal-to-noise ratio of
the phase fit --- that predicts success before the correction is applied ($r=0.85$
across $50$ networks). On a stationary limit cycle the correction merely matches the
trivial baseline of repeating the last shedding period; its value appears once the
flow ceases to repeat. On a wake driven by a slowly varying inflow, so that the
shedding frequency drifts and no period reproduces the next, repeating the last
period fails by over an order of magnitude while the phase correction which needs
only that the oscillation stay coherent, not exactly periodic --- still removes about
half of the rollout error. The correction is applied offline with no retraining.
\end{abstract}

\maketitle


%
\section{Introduction}
\label{sec:intro}

Direct numerical simulation of unsteady flows resolves every relevant
spatiotemporal scale and is, correspondingly, expensive. For applications that
require many evaluations of the same flow -- design optimisation, uncertainty
quantification, real-time control, digital twins -- the cost is prohibitive, and
reduced-order models offer an alternative: a low-dimensional surrogate,
constructed from data, that reproduces the dominant dynamics at a small fraction
of the cost.

The now-conventional construction proceeds in two stages. A spatial encoder --
classically proper orthogonal decomposition, and increasingly a convolutional
autoencoder -- compresses each flow field to a small number of latent
coordinates. A temporal model then advances those coordinates forward in time.
In its classical form the spatial encoder is proper orthogonal decomposition and
the reduced dynamics follow from Galerkin projection of the governing
equations\cite{berkooz1993pod, noack2003hierarchy, rowley2017modelreduction}.
Replacing the linear subspace with a convolutional autoencoder lifts the accuracy
ceiling that a linear projection imposes on advection-dominated
flows\cite{leecarlberg2020, murata2020mdcnn}, and coupling such an autoencoder to
a recurrent temporal model, most often an LSTM, has become a standard route to
fully data-driven surrogates of unsteady flows\cite{hasegawa2020bluff,
hasegawa2020cylinder, nakamura2021channel, REF-cae-lstm-rom}.

Whatever the temporal model, its use at inference is \emph{autoregressive}: having
been trained to predict one step ahead from true inputs, it must at deployment be
fed its own outputs. This mismatch between training and inference is the origin of
a failure mode so consistently reported that it is treated as an inherent property
of the approach. Small errors, introduced at each step, compound; over long
horizons the prediction degrades and eventually diverges. This behaviour is well documented: LSTM forecasts of chaotic systems lose accuracy
over long horizons and require an added stochastic term to remain on the
attractor\cite{vlachas2018lstm}; latent recurrent models of advection-dominated
flows accumulate error along the rollout\cite{REF-cae-lstm-rom}; and learned
time-steppers for partial differential equations show the same compounding of
single-step mistakes into eventual divergence\cite{brandstetter2022mppde,
sanchez2020simulate}.

The remedies proposed for this failure are numerous, and they share a common
premise. Training-side approaches unroll the model during training and penalise
the accumulated multi-step error, or inject noise so that the network learns to
recover from its own mistakes. This compounding is the \emph{exposure bias} of
autoregressive models, familiar from sequence generation\cite{bengio2015scheduled}
and increasingly noted in autoregressive weather and climate
emulation\cite{weyn2020, keisler2022}, where multi-step error growth is a central
obstacle. Remedies include the pushforward loss\cite{brandstetter2022mppde},
scheduled sampling\cite{bengio2015scheduled}, and training-noise
injection\cite{sanchez2020simulate} are representative. Architectural approaches constrain the latent dynamics
so that they cannot diverge -- by seeking coordinates in which the evolution is
linear, as in Koopman autoencoders\cite{lusch2018koopman, ottorowley2019}, or by
parameterising it spectrally, as in neural operators\cite{li2021fourier,
lu2021deeponet}. Post-hoc approaches leave the model frozen and
train a second network to predict and subtract the residual\cite{wang2020rnnclosure}.

The premise these share is that the rollout error is \emph{unstructured}: a
quantity to be suppressed, bounded, or regressed, but not one that has anything to
say. It is reported, almost without exception, as a scalar norm that grows with
the prediction horizon. The question of what the error \emph{is} -- what signal it
constitutes, what its temporal spectrum looks like, what component of the
prediction it corresponds to -- is not, to our knowledge, asked.

This paper asks it, for the case of periodic flows, and finds that the error is
highly structured.

Examining the rollout error of a convolutional-autoencoder--LSTM reduced-order
model of a two-dimensional cylinder wake as a time series rather than a magnitude,
we find that its power spectrum is not broadband but sharply peaked, and that the
peak lies at the vortex-shedding frequency of the flow. The same is true of a
square cylinder, which sheds by a different mechanism and at a distinctly
different frequency: there too the error peaks at that flow's own shedding
frequency. The error inherits the dominant frequency of the underlying physics.

Decomposing the error into amplitude and phase contributions establishes what this
means. Across the cases examined, $97$--$98\%$ of the error variance is
attributable to the \emph{phase} of the predicted oscillation and $2.0$--$2.3\%$ to
its amplitude, the predicted limit-cycle amplitude deviating from the true one by
only $0.15\%$ in root-mean-square. The network learns the geometry of the attractor
essentially exactly; what it fails to learn is the rate at which that attractor is
traversed. Its clock runs slightly slow, or slightly fast -- by a few thousandths
of a cycle over an entire rollout -- and that small, coherent timing error is the
whole of the long-horizon failure. This is the paper's central finding, and it
reframes what the rollout error of such a model \emph{is}: not an accumulation of
unstructured mistakes to be suppressed, but a single interpretable defect --- a drift
in the phase of an otherwise correctly-learned limit cycle.

The consequence of a phase error is amplified by the amplitude of the signal it
acts upon, which is why so small a defect produces an order-of-magnitude growth in
the error norm, and why its cause is invisible to the metrics conventionally
reported alongside it: a
network with $10^{-6}$ one-step validation error, correct to $0.15\%$ in amplitude,
and generating fields visually indistinguishable from the solver, is nonetheless
losing temporal alignment at a rate that will eventually render it useless.

Because the phase error accumulates linearly, it is described by a single number
per latent coordinate: the drift rate. This suggests a correction that acts on the
cause rather than the symptom. Rather than fitting a model to the error and
subtracting it, we estimate the drift rate over a short calibration window and
realign the phase of the prediction directly, leaving its amplitude untouched.

The resulting method has one parameter per latent coordinate, requires no
retraining, no auxiliary network, and no ground truth beyond the calibration
window. Where the linear phase model holds it reduces the extrapolation error by
$41$--$71\%$ on average across training runs, rising to $74$--$80\%$ on the runs
its own diagnostic accepts; in the single representative runs it removes
$72$--$83\%$ against $48$--$58\%$ for a correction fitted to the error itself,
and -- unlike that baseline -- its performance \emph{improves} with the prediction
horizon rather than degrading, which is the signature of a model that has captured
the mechanism rather than approximated its consequences. In physical space it
removes $91$--$98\%$ of the error that is removable at all, driving the
reconstruction to within a few per cent of the compression floor imposed by the autoencoder.

The method also reports its own applicability: the quality of the linear fit to the
accumulated phase determines whether a coherent drift exists to be corrected, and
this diagnostic reliably separates the cases in which the correction succeeds from
those in which it should not be attempted.

We stress, however, that the correction is a \emph{consequence} of the finding, not
its point. On a stationary limit cycle the simplest conceivable predictor --- copy
the last shedding period forward --- is already accurate, and there our correction
competes with it rather than decisively beating it. The distinction the finding
draws becomes operational only when the flow stops repeating: the correction
requires that the oscillation remain \emph{coherent}, so that an instantaneous phase
is defined, but not that it be exactly \emph{periodic}. To show that this is a real
distinction and not a semantic one, we include a wake driven by a slowly varying
inflow, in which the shedding frequency drifts continuously so that no cycle
reproduces the next. There, copying the last period fails by more than an order of
magnitude, while the phase correction --- operating on the still-coherent
instantaneous phase --- continues to remove roughly half of the rollout error. The
finding, that autoregressive latent rollout error is accumulated phase drift, thus
carries a practical corollary: the correction earns its place precisely where the
trivial baseline cannot be used.

The remainder of the paper is organised as follows.
Section~\ref{sec:methods} describes the flow configurations, the reduced-order
model, the amplitude--phase decomposition of the rollout error, and the proposed
correction. Section~\ref{sec:results} presents the spectral structure of the error,
its decomposition, the performance of the correction against baselines in both
latent and physical space, the conditions under which it applies, and its
behaviour on a non-stationary wake where periodic repetition is unavailable.
Section~\ref{sec:discussion} derives the observed structure analytically, explains
why a correction acting on the phase must outperform one acting on the error, and
sets out the limitations of the approach.
\section{Methodology}
\label{sec:methods}

\subsection{Flow configurations and direct numerical simulation}
\label{sec:methods:cfd}

Two canonical bluff-body configurations are considered: a circular cylinder and
a square cylinder, both in two dimensions. The two geometries are chosen because
they shed vortices by different mechanisms. Separation from the circular
cylinder occurs on a smooth surface and the separation points migrate with
Reynolds number; separation from the square cylinder is pinned at its sharp
upstream corners and is geometrically fixed. The two bodies also shed at
distinct Strouhal numbers (a factor of about $1.5$). A property of the reduced-order model that
persists across both configurations is therefore unlikely to be an artefact of a
single flow.

The incompressible Navier--Stokes equations,
\begin{equation}
    \frac{\partial \uvec}{\partial t}
    + (\uvec\cdot\nabla)\uvec
    = -\nabla p + \nu \nabla^{2}\uvec,
    \qquad
    \nabla\cdot\uvec = 0,
    \label{eq:ns}
\end{equation}
were solved with the finite-volume solver \texttt{pimpleFoam} (OpenFOAM v2512)
in laminar mode. Time integration used a second-order backward scheme with an
adaptive step limited to a maximum Courant number of $0.8$; spatial
discretisation was second-order central throughout.

The Reynolds number is $\Rey=U_\infty D/\nu$ and the Strouhal number
\begin{equation}
    \St = \frac{f_s D}{U_\infty},
    \label{eq:strouhal}
\end{equation}
where $f_s$ is the shedding frequency, $D$ the cylinder diameter or square side
length, and $U_\infty$ the free-stream velocity. In all cases $D = 0.12$ and
$U_\infty = 1.0$, so the Reynolds number is set through $\nu$ alone.

\paragraph{Circular cylinder.}
A body-fitted structured O-grid was used, with the flow simulated at
$\Rey = 300$--$1000$. Of these, $\Rey = 300$--$800$ provide the four cases of the
phase analysis and $\Rey = 1000$ enters as the negative case where the analysis
breaks down (Sec.~\ref{sec:results:breakdown}); the intermediate values carry the
seed and robustness studies. The
domain extends $5.06D$ upstream, $9.33D$
downstream, and $\pm5D$ laterally, with the cylinder centred at the origin; the
lateral boundaries are symmetry planes, the cylinder surface a no-slip wall, and
the outlet a fixed-pressure patch. The mesh comprises $34{,}764$ hexahedral
cells, with maximum non-orthogonality $43^{\circ}$ (mean $7^{\circ}$), maximum
skewness $0.39$, and maximum cell aspect ratio $2.4$.
Snapshots were written every $\Delta t = 0.03$ ($\approx 18$ per shedding cycle
for the circular cylinder);
the first 50 time units were discarded as transient, giving $N = 1334$ snapshots
per case.

\paragraph{Square cylinder.}
The square cylinder was simulated at $\Rey = 100$, within the regime in which the
wake is genuinely two-dimensional (the transition to three-dimensionality occurs
at $\Rey \approx 150$--$200$ for this geometry\cite{REF-square-2D3D}). The
configuration follows the established benchmark\cite{REF-square-benchmark}:
blockage ratio $5.6\%$ (domain height $17.9D$), $10D$ upstream and $30D$
downstream, the latter exceeding the $25D$ minimum required to preserve temporal
periodicity of the solution\cite{REF-outlet-distance}. The mesh was generated by
carving the square from a uniform Cartesian background grid; because the body
faces are grid-aligned the mesh is free of stair-stepping (maximum cell aspect
ratio $1.002$, maximum skewness $0.33$, mean non-orthogonality $2.1^{\circ}$;
$164{,}568$ cells, $48$ cells across the body face). Snapshots were written every
$\Delta t = 0.0438$ ($\approx 18$ per cycle), and the first 60 time units
discarded, giving $N = 1370$ snapshots.

\paragraph{Validation.}
For the square cylinder the lift coefficient was recorded throughout and its
spectrum used to obtain an independent measurement of the shedding frequency.
Over 37 cycles of the saturated limit cycle ($t = 90$--$120$) we obtain
$\St = 0.148$ and $\overline{C_D} = 1.502$, in agreement with published values
for this configuration ($\St \approx 0.137$--$0.15$,
$\overline{C_D} \approx 1.4$--$1.5$)\cite{REF-square-benchmark}. The lift
amplitude was constant to within $0.3\%$ across three consecutive windows,
confirming saturation before sampling began. For the circular cylinder the
measured Strouhal number rises monotonically from $0.213$ to $0.232$ over
$\Rey = 300$--$800$, consistent with the established $\St$--$\Rey$
relationship\cite{williamson1988, norberg2003}.

\paragraph{Grid and temporal convergence.}
\label{sec:methods:convergence}
Grid and timestep independence were verified for the circular cylinder at
$\Rey = 400$. The mesh was refined by a factor of $\sqrt{2}$ in each in-plane
direction (from $34{,}764$ to $69{,}296$ cells) and, independently, the timestep
was halved (reducing the maximum Courant number from $0.8$ to $0.4$); the
resulting changes in the mean drag coefficient and Strouhal number are summarised
in Table~\ref{tab:convergence}. The temporal discretisation is well converged, the
two quantities changing by at most $1.4\%$. The residual grid sensitivity
($\approx 2\%$ in $\overline{C_D}$ and $\St$) is common-mode to the reduced-order
analysis and does not enter the reported rollout error: the temporal model is
trained on, and evaluated against, simulation data from the \emph{same} mesh, so a
mesh-dependent offset in the absolute shedding frequency appears identically in the
training target and in the reference trajectory and cancels in the phase error,
which measures the model's detuning relative to its own training flow rather than
relative to the physically exact frequency. The reported phase drift is therefore
insensitive to the absolute grid convergence of the simulation.

We are nonetheless explicit about the limitations of the circular-cylinder
configuration, which is more constrained than the square-cylinder domain. The
outlet lies $9.3D$ downstream and the lateral boundaries at $\pm5D$, giving a
blockage of $10\%$; the square-cylinder domain, by contrast, extends $30D$
downstream at $5.6\%$ blockage. A domain of this size is adequate for a
statistically stationary limit cycle but not for high-accuracy absolute force or
frequency prediction, and we do not claim the latter. As a check, the computed
mean drag coefficient at $\Rey = 400$ is $\overline{C_D} = 1.59$, within the
scatter of published two-dimensional values at this Reynolds number
($\overline{C_D} \approx 1.5$--$1.6$); the Strouhal number
$\St = 0.219$ is consistent with the established $\St$--$\Rey$ relation for the
circular cylinder\cite{williamson1988, norberg2003} to within the $2\%$ grid
sensitivity quoted above. Because the
phase analysis depends only on the internal consistency of each simulation with
its own reduced-order model, and not on the absolute accuracy of these
quantities, we quote Strouhal numbers to no more than the two decimal places the
grid supports and attach the $2\%$ band wherever a comparison is made.

\begin{table}[htbp]
    \centering
    \caption{Grid and timestep convergence for the circular cylinder at
    $\Rey = 400$, evaluated over the saturated window $t = 50$--$90$.
    $\overline{C_D}$ is the time-averaged drag coefficient and $\St$ is obtained
    from the lift-coefficient spectrum. Halving the timestep changes both
    quantities by $\le 1.4\%$; doubling the mesh changes them by $\approx 2\%$.}
    \label{tab:convergence}
    \begin{tabular}{lccc}
        \toprule
        Case & Cells & $\overline{C_D}$ & $\St$ \\
        \midrule
        Production ($\mathrm{Co} = 0.8$)    & $34{,}764$ & $1.594$ & $0.219$ \\
        Fine mesh ($\mathrm{Co} = 0.8$)     & $69{,}296$ & $1.556$ & $0.224$ \\
        Half timestep ($\mathrm{Co} = 0.4$) & $34{,}764$ & $1.605$ & $0.222$ \\
        \midrule
        Grid change (production $\to$ fine)     & & $-2.4\%$ & $+2.2\%$ \\
        Timestep change (production $\to$ half) & & $+0.7\%$ & $+1.4\%$ \\
        \bottomrule
    \end{tabular}
\end{table}

\begin{figure}[htbp]
    \centering
     \includegraphics[width=\linewidth]{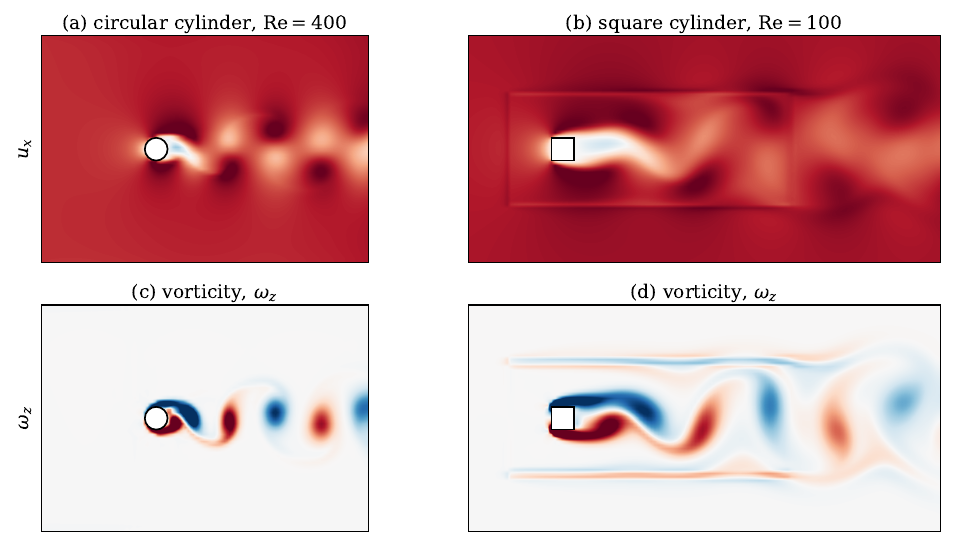}
    \caption{Flow configurations, both on the fully developed limit cycle.
    (a,b)~Instantaneous streamwise velocity $u_x$ for the circular cylinder at
    $\Rey=400$ and the square cylinder at $\Rey=100$. (c,d)~The corresponding
    spanwise vorticity $\omega_z$, showing the shed vortex street. Despite the
    different separation mechanisms, both flows settle onto a periodic limit
    cycle.}
    \label{fig:configurations}
\end{figure}

\subsection{Data preparation and reduced-order model}
\label{sec:methods:rom}

Each snapshot comprises the two velocity components and the pressure,
interpolated onto a uniform $128\times64$ Cartesian grid, giving a state
$\mathbf{q}_n \in \mathbb{R}^{3\times64\times128}$ ($24{,}576$ degrees of
freedom). For the square cylinder the extraction window was cropped to the body
and near wake, and grid points inside the solid masked to zero. Each channel was
standardised to zero mean and unit variance before training.

The reduced-order model follows the standard two-stage
construction\cite{REF-cae-lstm-rom}: a convolutional autoencoder compresses each
field to a latent vector, and a recurrent network advances that vector in time.

\begin{figure*}[htbp]
    \centering
    \resizebox{\linewidth}{!}{
%
%
\begin{tikzpicture}[
    font=\small,
    box/.style  = {draw, rounded corners=2pt, minimum height=8mm,
                   minimum width=15mm, align=center, thick},
    cae/.style  = {box, fill=blue!8,    draw=blue!55},
    lstm/.style = {box, fill=orange!12, draw=orange!70},
    corr/.style = {box, fill=teal!12,   draw=teal!65},
    dat/.style  = {align=center, font=\footnotesize, inner sep=1.5pt},
    ar/.style   = {-{Latex[length=1.8mm]}, thick},
    fb/.style   = {-{Latex[length=1.8mm]}, thick, dashed, red!75!black},
]

\node[dat]                       (q)    {$\mathbf{q}_n$};
\node[cae,  right=6mm of q]      (enc)  {encoder};
\node[dat,  right=6mm of enc]    (z)    {$\mathbf{z}_n$};
\node[lstm, right=6mm of z]      (net)  {LSTM};
\node[dat,  right=6mm of net]    (zh)   {$\hat{\mathbf{z}}_{n+1}$};

\draw[ar] (q)   -- (enc);
\draw[ar] (enc) -- (z);
\draw[ar] (z)   -- (net);
\draw[ar] (net) -- (zh);

\node[font=\scriptsize, above=0.5mm of q]  {$3\times64\times128$};
\node[font=\scriptsize, above=0.5mm of z]  {$n_z$};

\draw[fb, rounded corners=2pt]
      (zh.north) -- ++(0,5mm) -| (net.north);
\node[font=\scriptsize, red!75!black, align=center, above=5.5mm]
      at ($(net)!0.5!(zh)$)
      {at inference, the prediction becomes the next input};

\node[corr, below=14mm of zh]    (pc)   {phase\\correction};
\node[dat,  left=6mm of pc]      (zt)   {$\tilde{\mathbf{z}}_n$};
\node[cae,  left=6mm of zt]      (dec)  {decoder};
\node[dat,  left=6mm of dec]     (qt)   {$\tilde{\mathbf{q}}_n$};

\draw[ar] (zh) -- (pc);
\draw[ar] (pc) -- (zt);
\draw[ar] (zt) -- (dec);
\draw[ar] (dec) -- (qt);

\node[font=\scriptsize, teal!55!black, right=2mm of pc, align=left]
      {$\tilde{z}^{(d)}
        = \hat{A}^{(d)}\cos\!\big(\hat{\theta}^{(d)} - \omega_d\,t\big)
          + \overline{\hat{z}^{(d)}}$};

\node[font=\scriptsize, align=center, below=1mm of pc]
      {\emph{post hoc} --- never fed back};

\end{tikzpicture}}
    \caption{The reduced-order model and the proposed correction. A convolutional
    autoencoder maps the flow field to a latent vector $\zvec$; an LSTM advances
    $\zvec$ autoregressively. The correction acts on the predicted latent
    trajectory \emph{after} the rollout and is never fed back into the network.}
    \label{fig:pipeline}
\end{figure*}
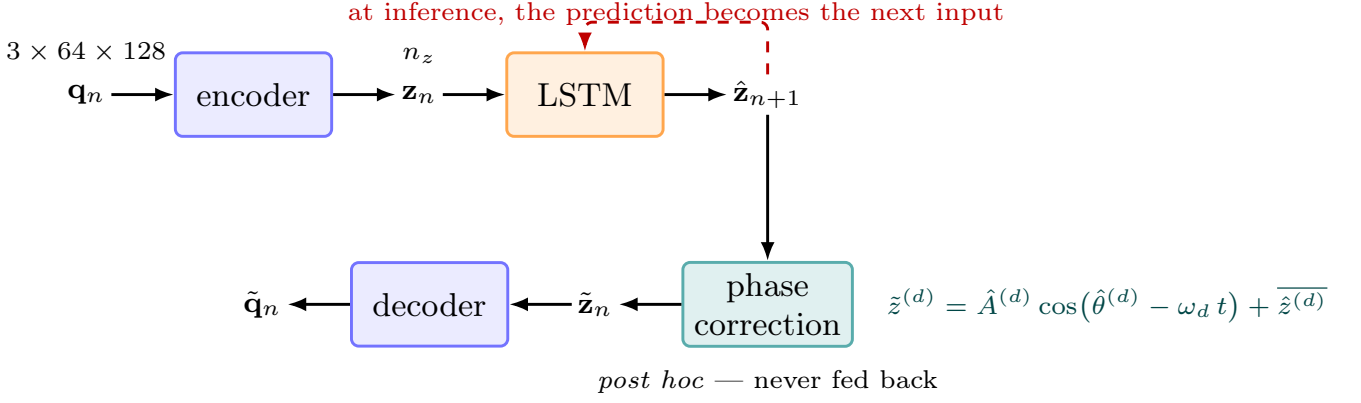

\subsubsection{Convolutional autoencoder}

The encoder comprises four convolutional layers with $3\times3$ kernels, the
latter three with stride two, reducing the $64\times128$ field to $8\times16$
while the channel count grows from $3$ to $64$; a linear layer then maps the
resulting $8192$ activations to the latent vector $\zvec\in\mathbb{R}^{n_z}$. The
decoder mirrors this using transposed convolutions. ReLU activations are used
throughout except at the bottleneck and output, both linear: the latent
coordinates and the reconstructed fields must be free to take negative values.

Training minimised mean-squared reconstruction error (Adam, $10^{-3}$, 300
epochs, best-validation checkpoint retained). Latent dimensions $n_z = 4$ and
$n_z = 8$ were used.

Because compression is lossy, the autoencoder imposes an irreducible floor on any
latent-space prediction. We write $\varepsilon_{\mathrm{CAE}}$ for the field-space
RMSE obtained by encoding and immediately decoding the \emph{true} snapshots. No
temporal model, however accurate, can produce a reconstruction better than this
floor.

\subsubsection{Temporal model}

A two-layer LSTM\cite{REF-hochreiter} with 64 hidden units advances the latent
state. Given a window of $\tau=20$ latent vectors it predicts the next,
\begin{equation}
    \hat{\zvec}_{n+1} = \mathcal{F}\!\left(
        \zvec_{n-\tau+1},\dots,\zvec_{n}\right),
    \label{eq:lstm}
\end{equation}
trained on the one-step mean-squared error under teacher forcing (Adam, 500
epochs, gradient-norm clipping at $1.0$). At inference the network is run
\emph{autoregressively}, the window advancing on its own outputs:
\begin{equation}
    \hat{\zvec}_{n+1} = \mathcal{F}\!\left(
        \hat{\zvec}_{n-\tau+1},\dots,\hat{\zvec}_{n}\right).
    \label{eq:rollout}
\end{equation}
This train--inference mismatch is the origin of the rollout error studied here.
The snapshots were split in time: $52\%$ training, $8\%$ validation, and the
remaining $\approx40\%$ reserved as the rollout horizon.

\subsection{Characterising the rollout error}
\label{sec:methods:error}

The rollout error is
\begin{equation}
    \evec_n = \hat{\zvec}_n - \zvec_n \in \mathbb{R}^{n_z}.
    \label{eq:error}
\end{equation}
Conventionally this enters an analysis only through its magnitude
$\|\evec_n\|$. We instead treat each component as a time series and examine it in
two ways.

\subsubsection{Spectral content}

The power spectrum of the error in latent coordinate $d$ is
\begin{equation}
    S^{(d)}(f) = \left| \sum_{n=0}^{N-1}
        \left( e_n^{(d)} - \bar{e}^{(d)} \right)
        \mathrm{e}^{-\mathrm{i}2\pi f n \Delta t}\right|^{2}.
    \label{eq:error-spectrum}
\end{equation}
This is an unnormalised periodogram of the mean-removed series; no window or
detrending is applied, so the quoted peak-to-floor ratios are unaffected by
window leakage.

\subsubsection{Amplitude--phase decomposition}
\label{sec:methods:decomp}

To determine \emph{what} the error is, rather than merely at which frequency it
appears, both the true and the predicted latent trajectories are written in
analytic form via the Hilbert transform $\mathcal{H}$. For a real signal $x(t)$
the analytic signal is $x + \mathrm{i}\mathcal{H}[x]$, whose modulus and argument
give an instantaneous amplitude and an unwrapped instantaneous phase. The
analytic signal has known end artefacts, and we verified their influence directly:
discarding the first $5$--$20$ samples of the calibration window changes the
fitted drift rate by $4$--$14\%$. We report the untrimmed estimate throughout and
treat this range as the systematic uncertainty on $\omega_d$; we do \emph{not}
select the trim by extrapolation performance, which would be circular given that
the extrapolation region is also the scoring region. A phase estimator that avoids
the Hilbert end effect entirely --- a band-pass around $f_s$ before the transform,
or a geometric phase from a pair of latent coordinates --- would remove this
ambiguity, and we identify it as the natural robustness check
(Sec.~\ref{sec:discussion:limits}). This
representation is well-posed because each latent coordinate is narrowband. Taking
the unwindowed periodogram of the mean-removed signal and excluding the zero
frequency, $98$--$99\%$ of the spectral energy lies within $\pm10\%$ of the
shedding frequency; the residual is distributed between the second harmonic and a
broadband floor, and is consistent with the in-fit residual of the two-harmonic
model of Eq.~\eqref{eq:harmonic} (Sec.~\ref{sec:results:dmd}). We state the
criterion explicitly because the figure is sensitive to it: applying a Hann window
before the transform raises the apparent in-band fraction to $99.9\%$ by
broadening the peak, and narrowing the band to $\pm5\%$ lowers it to $96\%$. At
$98$--$99\%$ the signal is comfortably narrowband in the sense required for the
analytic signal to carry an unambiguous instantaneous phase, while remaining far
from the purely harmonic limit --- a distinction that matters in
Sec.~\ref{sec:results:dmd}. Writing
\begin{equation}
    z^{(d)}(t)       = A^{(d)}(t)\cos\theta^{(d)}(t),
    \qquad
    \hat{z}^{(d)}(t) = \hat{A}^{(d)}(t)\cos\hat{\theta}^{(d)}(t),
    \label{eq:analytic-form}
\end{equation}
the error separates into the contribution that would arise from the amplitude
alone being wrong,
\begin{equation}
    e_{A}^{(d)}(t) = \bigl(\hat{A}^{(d)} - A^{(d)}\bigr)\cos\theta^{(d)},
    \label{eq:e-amp}
\end{equation}
and that which would arise from the phase alone being wrong,
\begin{equation}
    e_{\theta}^{(d)}(t)
        = A^{(d)}\bigl(\cos\hat{\theta}^{(d)} - \cos\theta^{(d)}\bigr),
    \label{eq:e-phase}
\end{equation}
together with a cross term $e_{A\theta}^{(d)}$ that measures their interaction.
Amplitude fidelity is quantified by the root-mean-square deviation
$\langle (\hat{A}^{(d)}-A^{(d)})^2 \rangle^{1/2}$ over the whole rollout, relative
to the mean true amplitude, rather than by the ratio of terminal amplitudes: the
latter is a single-point statistic that understates the envelope fluctuation.
The variance of each contribution, as a fraction of the total error variance,
identifies which mechanism dominates. The decomposition is not orthogonal, so
these shares need not sum exactly to unity; the magnitude of the cross term is
itself the relevant diagnostic. In every well-posed case it is negligible
($<10^{-5}$ of the total variance), so the amplitude and phase shares partition
the error to within a fraction of a percent, whereas where it grows to $O(1)$ the
representation has broken down (Sec.~\ref{sec:results:decomp}). This decomposition
is the observation on which the correction is built.

The accumulated phase error is
\begin{equation}
    \phi^{(d)}(t) = \hat{\theta}^{(d)}(t) - \theta^{(d)}(t),
    \label{eq:phi}
\end{equation}
referenced to zero at the start of the rollout.

\subsection{Correction by phase realignment}
\label{sec:methods:phasecorr}

Anticipating the results of Secs.~\ref{sec:results:decomp}
and~\ref{sec:results:linear} -- that the error is almost entirely a phase error,
and that the phase drifts linearly -- we correct the prediction by realigning its
phase rather than by modelling the error it produces.

The phase-drift rate $\omega_d$ is obtained by linear regression of
$\phi^{(d)}(t)$ over a calibration window of $N_{\mathrm{cal}}$ steps, during
which ground truth is assumed available. The corrected trajectory is then
reconstructed directly from the prediction's own analytic representation:
\begin{equation}
    \tilde{z}^{(d)}(t)
        = \hat{A}^{(d)}(t)\,
          \cos\!\left( \hat{\theta}^{(d)}(t) - \omega_d t \right)
          + \overline{\hat{z}^{(d)}} .
    \label{eq:phase-correction}
\end{equation}
Here $\overline{\hat{z}^{(d)}}$ denotes the \emph{time-mean} of the predicted
coordinate, not the predicted signal itself: the analytic signal is formed from
the mean-removed trajectory, so $\hat{A}^{(d)}\cos\hat{\theta}^{(d)}$ reconstructs
only the oscillatory part and the mean must be restored. The predicted amplitude
is retained unaltered; only the phase is adjusted. The
method has a \emph{single} parameter per latent coordinate, requires no amplitude
model, no growth envelope and no saturation cap, and -- because $\phi$ is linear
-- its extrapolation beyond the calibration window is exact rather than
approximate.

Two properties of Eq.~\eqref{eq:phase-correction} should be stated plainly. First,
$\hat{A}^{(d)}$ and $\hat{\theta}^{(d)}$ are obtained from the analytic signal of
the \emph{complete} predicted trajectory, so although no ground truth is used
beyond the calibration window, the correction is applied off-line and is not
causal; a streaming implementation would require a one-sided or sliding-window
estimate of the analytic signal, which we have not tested. Second, the regression
of $\phi^{(d)}$ yields both a slope and an intercept, and only the slope is
applied. The intercept represents a constant phase offset already present at the
start of the rollout rather than an accumulating error; removing it would alter
the initial condition, which we prefer to leave untouched, and it is small
($\lesssim 5\times10^{-3}$\,rad) compared with the drift accumulated over the
horizon.

Equation~\eqref{eq:phase-correction} is evaluated pointwise and involves no
resampling. An equivalent formulation as a time warp, in which the prediction is
re-evaluated at shifted times, fails in practice: the required shift is a small
fraction of one sampling interval, and interpolating an oscillatory signal at
sub-sample offsets introduces more error than the correction removes.

\paragraph{Applicability.}
The coefficient of determination $R^2$ of the linear fit to $\phi(t)$ over the
calibration window measures whether the model on which the correction rests
actually holds. Where $R^2$ is small, either there is no coherent phase drift to
remove or it cannot be resolved from the available calibration data; in either
case the fitted $\omega$ is dominated by noise and the correction should not be
applied. Because $\phi$ is small and bounded, $R^2$ measures principally the
signal-to-noise ratio of the phase estimate rather than any departure from
linearity; the threshold used below is therefore an empirical SNR criterion. We
report $R^2$ -- computed per latent coordinate over the calibration window and
quoted as the coordinate mean -- alongside every result, and show in
Sec.~\ref{sec:results:applicability} that it functions as a reliable
self-diagnostic.

\subsubsection{Baseline: direct fitting of the error}
\label{sec:methods:sinusoid}

For comparison we implement the correction that follows from the spectral
observation alone, without the decomposition: since the error is periodic at
$f_s$ with a growing envelope, fit a growing sinusoid to it and subtract. Per
latent coordinate,
\begin{equation}
    c^{(d)}(t) = \mathrm{clip}\!\left[
        A_d\, \mathrm{e}^{\gamma_d t}\,
        \sin\!\left( \hat{\theta}^{(d)}(t) + \psi_d \right), \;\pm\kappa_d
    \right],
    \label{eq:sinusoid}
\end{equation}
with $(A_d,\gamma_d,\psi_d)$ fitted by nonlinear least squares on the same
calibration window, $\gamma_d \ge 0$, and the cap
$\kappa_d = 1.5\max_{n\le N_{\mathrm{cal}}}|e_n^{(d)}|$ preventing the exponential
from diverging beyond it. The corrected trajectory is
$\tilde{\zvec}_n = \hat{\zvec}_n - \mathbf{c}_n$. This baseline requires three
parameters per coordinate and a saturation cap; the comparison in
Sec.~\ref{sec:results:correction} quantifies what is gained by correcting the
cause rather than the symptom.

Both corrections are applied \emph{post hoc}. Neither is fed back into the LSTM:
the network's inputs during rollout are always its own raw predictions.
Perturbing the input window with corrected values moves the network away from the
distribution on which it was trained and, in our experiments, degraded the
prediction severely.

\subsection{Configurations examined}
\label{sec:methods:matrix}

To make the case matrix explicit: the circular cylinder was simulated at
$\Rey = 300$, $400$, $500$, $600$, $700$, $800$ and $1000$, and the square
cylinder at $\Rey = 100$. Reduced-order models were trained at latent dimension
$n_z = 4$ for every case and additionally at $n_z = 8$ for $\Rey = 800$ and
$\Rey = 100$. The phase analysis is reported for four configurations --- circular
$\Rey = 300$, $400$ and $800$ and square $\Rey = 100$ --- chosen to span both
geometries, a factor of $2.7$ in Reynolds number and both latent dimensions.
Three further configurations appear in the text where they isolate a specific
effect: $\Rey = 800$ at $n_z = 4$ and $\Rey = 100$ at $n_z = 8$ separate
latent-dimension effects from Reynolds-number effects
(Sec.~\ref{sec:results:applicability}), and $\Rey = 500$ carries the seed study of
Sec.~\ref{sec:results:seeds}, having been chosen for that purpose because its
drift magnitude is intermediate. $\Rey = 1000$ is analysed only to establish the
limits of the method. Simulations at $\Rey = 600$ and $700$ were also
performed; they behave as the neighbouring cases do and are not analysed
separately.

One further configuration is non-stationary. To test the correction on a flow that
is coherent but not periodic (Sec.~\ref{sec:results:modulated}), the
$\Rey = 400$ circular-cylinder inflow is modulated in time,
$U(t) = U_0\,[1 + \varepsilon\sin(2\pi f_m t)]$ with $\varepsilon = 0.05$ and
$f_m$ chosen so that one modulation period spans about fifteen shedding cycles.
The instantaneous Reynolds number then drifts between $380$ and $420$; the shedding
remains coherent at every instant while its frequency tracks the inflow, so that no
shedding period reproduces the next. All numerics are otherwise identical to the
stationary $\Rey = 400$ case, and the reduced-order model is trained on this flow
without modification.

\subsection{Linear-propagator baselines}
\label{sec:methods:dmd}

The latent trajectories analysed here are narrowband and of near-constant
amplitude, which raises a natural objection: if the latent dynamics are so nearly
harmonic, a linear propagator should suffice, and a recurrent network is
unnecessary. We therefore compare against two linear alternatives that share the
same autoencoder and differ only in how the latent state is advanced.

\paragraph{Dynamic mode decomposition.}
Dynamic mode decomposition (DMD)\cite{schmid2010dmd, tu2014dmd} seeks the single
linear operator that best advances the state by one step. Assembling the latent
snapshots into the time-shifted matrices
$\mathbf{Z} = [\zvec_0\;\cdots\;\zvec_{m-1}]$ and
$\mathbf{Z}' = [\zvec_1\;\cdots\;\zvec_{m}]$, the operator is the least-squares
solution
\begin{equation}
    \mathbf{A} \;=\; \arg\min_{\mathbf{A}}
        \bigl\| \mathbf{Z}' - \mathbf{A}\mathbf{Z} \bigr\|_F
    \;=\; \mathbf{Z}'\mathbf{Z}^{+},
    \label{eq:dmd}
\end{equation}
with $\mathbf{Z}^{+}$ the Moore--Penrose pseudoinverse. Prediction is then
repeated application, $\zvec_n = \mathbf{A}^n \zvec_0$. Because the latent
dimension is small ($n_z \le 8$) the pseudoinverse is formed directly and
Eq.~\eqref{eq:dmd} is exact; the rank-truncated formulation
of Ref.~\onlinecite{tu2014dmd} is unnecessary here.

The eigendecomposition $\mathbf{A}\boldsymbol{\varphi}_j =
\lambda_j\boldsymbol{\varphi}_j$ separates the dynamics into modes evolving
independently as $\lambda_j^{\,n}$. Each eigenvalue carries two pieces of
physical information: $|\lambda_j|$ is a growth or decay rate, with
$|\lambda_j| = 1$ denoting a neutrally stable oscillation, and
$\arg\lambda_j/(2\pi\Delta t)$ is the mode's frequency. A limit cycle is
represented by a conjugate pair on the unit circle at the shedding frequency, so
the spectrum provides a direct check that the operator has been fitted correctly.
DMD is the finite-dimensional realisation of Koopman spectral
analysis\cite{lusch2018koopman}, and has been applied to the cylinder wake
specifically\cite{bagheri2013cylinder}; combining it with a nonlinear encoder is
therefore a principled construction rather than a deliberately weak reference.

Three variants are evaluated, and the strongest is reported in every comparison:
$\mathbf{A}$ fitted on the training window; the same operator with its
eigenvalues projected onto the unit circle, $\lambda_j \mapsto
\lambda_j/|\lambda_j|$, which removes spurious growth and decay; and
$\mathbf{A}$ refitted on the training window \emph{together with} the calibration
window, so that the linear baseline receives exactly the ground-truth data the
phase correction uses.

\paragraph{Harmonic model.}
The second baseline abandons the state-space form and fits each latent coordinate
directly as a truncated Fourier series at its dominant frequency,
\begin{equation}
    z^{(d)}(t) \;\approx\; c_0^{(d)}
    + \sum_{h=1}^{2}\Bigl[
        a_h^{(d)}\cos\bigl(2\pi h f_0^{(d)} t\bigr)
      + b_h^{(d)}\sin\bigl(2\pi h f_0^{(d)} t\bigr) \Bigr],
    \label{eq:harmonic}
\end{equation}
with $f_0^{(d)}$ obtained from the periodogram peak refined by parabolic
interpolation and the coefficients by linear least squares on all data preceding
the rollout. Extrapolation is evaluation of Eq.~\eqref{eq:harmonic} at later
times. This model has constant amplitude by construction, so its only available
failure mode is phase error --- a property we exploit in
Sec.~\ref{sec:results:dmd}.

\subsection{Training and implementation details}
\label{sec:methods:training}

The autoencoder and LSTM are trained with Adam at an initial learning rate of
$10^{-3}$, reduced on plateau, for up to $300$ and $500$ epochs respectively with
early stopping on a held-out validation split. Within each seed ensemble a fresh
autoencoder is trained per seed, so latent coordinates are not shared across
seeds. In total this study trains $50$ networks for the latent-space efficacy
ensemble, a separate $40$ for the field-space ensemble, $40$ for the
drift-predictor null of Sec.~\ref{sec:results:seeds}, $40$ for the
noise-injection test, and single networks for the $n_z=16$ and intermediate-$\Rey$
cases. Field-space errors are computed over the full cropped domain; for the
square cylinder the solid region is a small fraction of the crop and is retained
in the root-mean-square denominators, so the reported $\varepsilon_{\mathrm{CAE}}$
and $G$ are conservative. Computations use PyTorch~2.x with CUDA~12; the analysis
code and case files are available at the repository below. The analysis and
figure-generation scripts were developed with assistance from a large language
model (Claude Opus~4.8, Anthropic) and were independently verified against the
simulation data by the authors.

\subsection{The correction-efficacy ratio $\rho$}
\label{sec:methods:rhodef}

Whether a latent-space correction improves the \emph{reconstructed field} depends
on whether the temporal prediction error is large compared with the compression
error the autoencoder imposes regardless. We define
\begin{equation}
    \rho \;=\;
    \frac{\sqrt{\,\varepsilon_{\mathrm{unc}}^{2}
                - \varepsilon_{\mathrm{CAE}}^{2}\,}}
         {\varepsilon_{\mathrm{CAE}}},
    \label{eq:rho}
\end{equation}
where $\varepsilon_{\mathrm{unc}}$ is the field-space RMSE of the uncorrected
rollout. The numerator isolates the part of the field error attributable to the
temporal model, under the assumption that compression and temporal errors combine
in quadrature, $\varepsilon_{\mathrm{unc}}^{2} \approx
\varepsilon_{\mathrm{CAE}}^{2} + \varepsilon_{\mathrm{temporal}}^{2}$. Strictly
this requires the field-space image of the temporal error to be orthogonal to the
compression error, which is not guaranteed through a nonlinear decoder; we
therefore treat $\rho$ as an interpretable diagnostic rather than an exact
decomposition. In every case reported here the radicand remained positive
throughout the extrapolation window, so $\rho$ is well defined; a negative
radicand would indicate that the assumption has failed and that $\rho$ should not
be used. Thus $\rho \gg 1$ indicates a prediction-limited regime and
$\rho \ll 1$ a compression-limited one.

$\rho$ and $R^2$ answer different questions and both are required. $R^2$ asks
whether the phase-drift model is valid, and hence whether the correction can be
constructed at all. $\rho$ asks whether the resulting latent-space improvement
will be visible after decoding, or will be masked by the compression floor.

\subsection{Reported metrics}
\label{sec:methods:metrics}

Performance is quantified over the \emph{extrapolation region} only,
$n > N_{\mathrm{cal}}$; scoring on the calibration window would be circular. In
latent space we report
\begin{equation}
    I = \left(1 -
        \frac{\langle \|\tilde{\zvec}_n - \zvec_n\| \rangle}
             {\langle \|\hat{\zvec}_n - \zvec_n\| \rangle}\right)\times 100\%.
    \label{eq:improvement}
\end{equation}

In field space the corresponding relative reduction is bounded above by the
autoencoder floor, which accounts for the majority of the total field error and
which no temporal correction can address. We therefore also report the fraction
of the \emph{correctable} error that is removed,
\begin{equation}
    G = \left(1 -
        \frac{\varepsilon_{\mathrm{cor}} - \varepsilon_{\mathrm{CAE}}}
             {\varepsilon_{\mathrm{unc}} - \varepsilon_{\mathrm{CAE}}}\right)
        \times 100\%,
    \label{eq:gap-closed}
\end{equation}
that is, the proportion of the excess of the uncorrected reconstruction over the
compression floor that the correction eliminates. $G = 100\%$ would mean the
corrected reconstruction has reached the compression limit exactly.
\section{Results}
\label{sec:results}

\subsection{The rollout error is a coherent oscillation}
\label{sec:results:spectrum}

\begin{figure}[htbp]
    \centering
    \includegraphics[width=\linewidth]{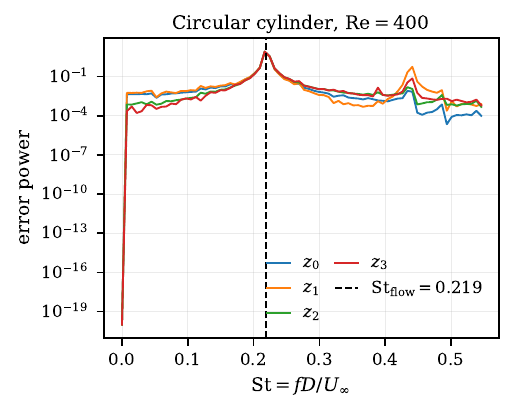}
    \caption{Power spectrum of the latent rollout error, circular cylinder,
    $\Rey = 400$, $n_z = 4$. All four latent coordinates peak at $\St = 0.219$,
    coinciding with the vortex-shedding frequency. The error is a coherent
    oscillation, not broadband noise.}
    \label{fig:spectrum-cyl}
\end{figure}

Figure~\ref{fig:spectrum-cyl} shows the power spectrum of the latent rollout
error, Eq.~\eqref{eq:error-spectrum}, for the circular cylinder at $\Rey = 400$
with $n_z = 4$. The spectra of all four latent coordinates collapse onto a single
sharp peak, three to four orders of magnitude above the surrounding broadband
floor, located at $\St = 0.219$ -- the shedding Strouhal number of the flow.

The error is therefore not unstructured. Were it the accumulation of independent
small mistakes, as it is usually implicitly treated, its spectrum would be
broadband. Instead essentially all of its power lies at a single frequency, and
that frequency is the one at which the physical flow oscillates.

\subsection{The error peak follows the shedding frequency of each flow}
\label{sec:results:generality}

A peak at the shedding frequency in one flow is suggestive but not conclusive: it
could reflect a property of the network, of the training, or of the particular
geometry. We therefore repeat the analysis on the square cylinder, which sheds by
a different mechanism and at a distinctly different frequency ($\St = 0.148$
versus $0.213$--$0.232$ for the circular cylinder).

\begin{figure}[htbp]
    \centering
    \includegraphics[width=\linewidth]{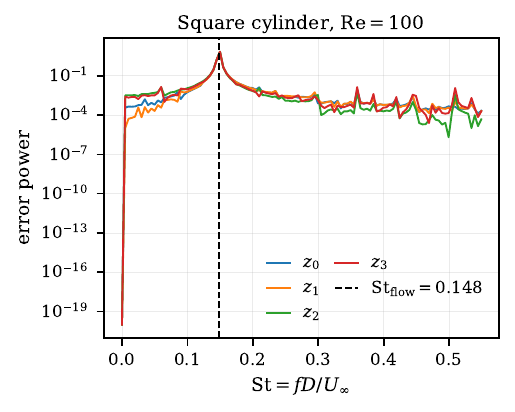}
    \caption{Power spectrum of the latent rollout error, square cylinder,
    $\Rey = 100$, $n_z = 4$. The peak lies at $\St = 0.149$, matching this flow's
    measured shedding frequency ($\St = 0.148$) and differing markedly from that
    of the circular cylinder.}
    \label{fig:spectrum-sq}
\end{figure}

Figure~\ref{fig:spectrum-sq} shows the result. The error spectrum is again
dominated by a single sharp peak. With a record of $16$ time units the spectral
resolution is $\Delta\St \approx 0.008$, so we do not over-read the peak location;
to that resolution it coincides with the square cylinder's shedding frequency
($\St \approx 0.15$) and lies nowhere near the circular cylinder's
($\St \approx 0.22$). The rollout error
inherits whatever dominant frequency the underlying flow possesses.

\subsection{The error is a phase error}
\label{sec:results:decomp}

That the error oscillates at $f_s$ establishes that it is structured, but not
what the structure \emph{is}. Two quite different failures would both produce a
peak at $f_s$: the network might reproduce the correct oscillation with the wrong
amplitude, or the correct amplitude with the wrong timing. The
amplitude--phase decomposition of Sec.~\ref{sec:methods:decomp} distinguishes
them.

The answer is unambiguous. Table~\ref{tab:decomposition} reports the variance
share of each contribution. Across the four representative runs the phase term
accounts for $97$--$98\%$ of the error variance and the amplitude term for
$2.0$--$2.3\%$ (Table~\ref{tab:decomposition}); the wider range $95$--$98\%$
quoted elsewhere refers to the seed ensemble of Sec.~\ref{sec:results:seeds},
whose lower tail extends further. Amplitude fidelity is quantified by the
root-mean-square deviation of the predicted from the true limit-cycle amplitude,
which is $0.12$--$0.14\%$ of the mean amplitude; we do not use the ratio of
terminal amplitudes, a single-point statistic that understates the envelope
fluctuation.

The decomposition is not orthogonal, but in every case in Table~\ref{tab:decomposition}
the cross term is negligible ($<10^{-5}$ of the total variance), so the amplitude
and phase shares account for the error to within $1\%$ (the residual being the
non-orthogonality of the decomposition, largest at $\Rey=300$). This is the criterion by
which $\Rey=1000$ is judged to lie outside the method's scope: there the cross
term rises to $O(1)$ and the phase share exceeds unity, a genuine breakdown of the
representation rather than the marginal excess over $100\%$ that near-orthogonality
alone produces.

\begin{figure}[htbp]
    \centering
    \includegraphics[width=\linewidth]{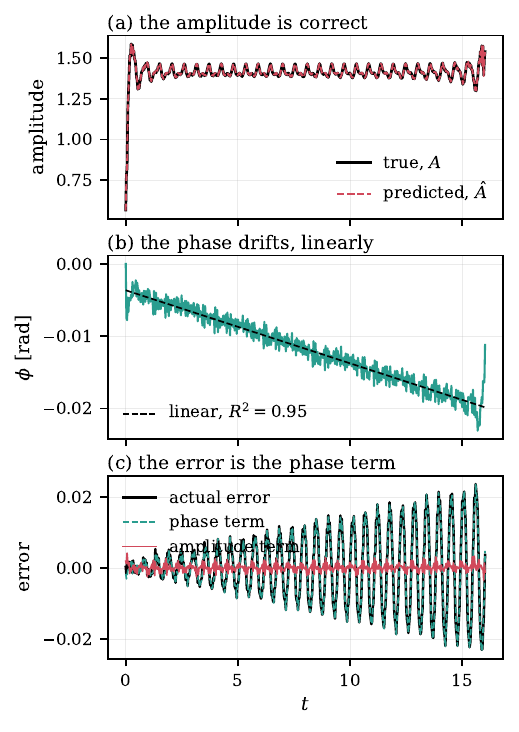}
    \caption{Amplitude--phase decomposition of the rollout error (circular
    cylinder, $\Rey=400$, latent coordinate $z_0$). (a) The predicted and true
    amplitude envelopes are indistinguishable. (b) The accumulated phase error
    grows steadily. (c) The error reconstructed from the phase term alone
    reproduces the actual error.}
    \label{fig:decomposition}
\end{figure}

The network learns the shape of the limit cycle essentially perfectly. What it
gets wrong is the timing. Figure~\ref{fig:decomposition} makes this visible: the
predicted and true amplitude envelopes lie on top of one another, while the phase
difference between prediction and truth grows steadily; and the error
reconstructed from the phase term alone is indistinguishable from the actual
error.

\begin{table}[htbp]
    \centering
    \caption{Variance share of the amplitude and phase contributions to the
    rollout error (single representative run per case, $N_{\mathrm{cal}}=150$),
    and the root-mean-square deviation of the predicted limit-cycle amplitude from
    the true amplitude over the whole rollout, relative to the mean true
    amplitude. The error is overwhelmingly a phase error.}
    \label{tab:decomposition}
    \begin{tabular}{llcccc}
        \toprule
        & & & \multicolumn{2}{c}{variance share} & \\
        \cmidrule(lr){4-5}
        Geometry & $\Rey$ & $n_z$ & phase & amplitude & RMS $|\hat A - A|/A$ \\
        \midrule
        Circular & 300 & 4 & $97.0\%$ & $2.3\%$ & $0.12\%$ \\
        Circular & 400 & 4 & $98.2\%$ & $2.0\%$ & $0.14\%$ \\
        Circular & 800 & 8 & $97.9\%$ & $2.3\%$ & $0.13\%$ \\
        Square   & 100 & 4 & $97.5\%$ & $2.0\%$ & $0.12\%$ \\
        \bottomrule
    \end{tabular}
\end{table}

\subsection{The phase drifts linearly}
\label{sec:results:linear}

If the error is a phase error, the quantity that governs it is the accumulated
phase difference $\phi(t)$, Eq.~\eqref{eq:phi}. Figure~\ref{fig:decomposition}(b)
shows that $\phi$ grows linearly. Over the full rollout it reaches only
$\approx 0.02$--$0.03$\,rad -- some $0.003$--$0.005$ of a single cycle -- yet this
suffices to produce an order-of-magnitude growth in the error, because the error
scales with $\phi$ and the latent oscillation has $O(1)$ amplitude.

Linearity of $\phi$ has a strong consequence: the entire error is characterised
by a \emph{single} number per latent coordinate, the drift rate
$\omega_d = \mathrm{d}\phi^{(d)}/\mathrm{d}t$, and its extrapolation beyond the
calibration window is exact rather than approximate. This is the basis of the
correction of Sec.~\ref{sec:methods:phasecorr}.

The quality of the linear fit varies between cases, and this variation is
informative rather than incidental; we return to it in
Sec.~\ref{sec:results:applicability}.

\subsection{Correcting the cause outperforms fitting the symptom}
\label{sec:results:correction}

Table~\ref{tab:correction} compares the phase realignment of
Eq.~\eqref{eq:phase-correction} against the sinusoid baseline of
Eq.~\eqref{eq:sinusoid}, over the extrapolation region and with an identical
calibration budget of $N_{\mathrm{cal}} = 150$ steps ($\approx 8$ shedding
cycles).

\begin{figure}[htbp]
    \centering
    \includegraphics[width=\linewidth]{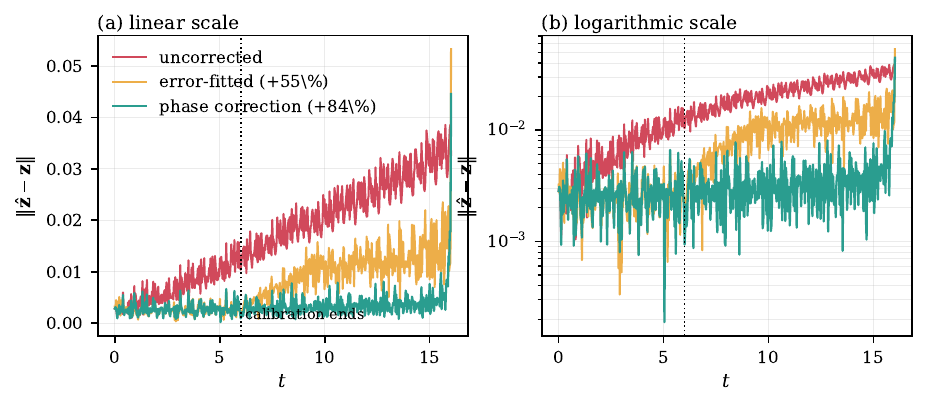}
    \caption{Latent-space error over the rollout, circular cylinder, $\Rey=400$.
    The correction is calibrated on the first 150 steps (dashed line) and
    evaluated thereafter. The phase correction holds the error nearly flat across
    the extrapolation horizon.}
    \label{fig:correction}
\end{figure}

Where the drift is resolvable above the phase-estimation jitter, the phase
correction reduces the latent
extrapolation error by $72$--$83\%$ in the single representative runs of
Fig.~\ref{fig:correction}, against $48$--$58\%$ for the sinusoid -- an improvement
of $21$--$29$ percentage points (single runs; $20.5$--$28.5$ computed from Table~\ref{tab:correction}), achieved with one parameter per
latent coordinate rather than three, and with no growth envelope and no
saturation cap.

The behaviour with horizon is the more telling comparison. Table~\ref{tab:horizon}
reports the improvement of each method as the evaluation window is extended.
The sinusoid degrades: its envelope model is fitted on the calibration window and
becomes progressively less valid thereafter, and its saturation cap eventually
truncates a correction that the true error has outgrown. The phase correction
does the opposite -- it \emph{improves} with horizon, from $75\%$ at 96 steps
beyond calibration to $78\%$ at 385. This is the signature of a model that
captures the actual mechanism: a constant-rate clock error extrapolates exactly,
so the longer the rollout, the larger the fraction of the total error it accounts
for.

\begin{table*}[htbp]
    \centering
    \caption{Correction efficacy over the extrapolation region (percentage
    reduction in latent error). All entries use the same calibration window
    $N_{\mathrm{cal}}=150$. The single-run columns are one representative training
    run per case, including the fitted drift rate $\omega_d$ (rad per time unit,
    coordinate-mean; per-coordinate values agree to within $4$--$13\%$, so the
    realignment is effectively single-parameter). The ensemble columns are
    mean\,$\pm$\,standard deviation over $n$ independently trained networks,
    unconditionally and restricted to networks for which the drift is resolvable
    above the phase-estimation jitter ($R^2>0.3$), the prospective criterion of
    Sec.~\ref{sec:results:applicability}. This latent-space ensemble is trained
    independently of the field-space ensemble of Table~\ref{tab:field}; the two
    share seed indices but not weights, so their accepted counts differ. Conditioning on $R^2$ --- which uses no
    information beyond the calibration window --- narrows the spread from
    $\pm25$--$42$ to $\pm2$--$8$ percentage points and brings all four cases into
    the range $74$--$80\%$. The decision rule is prospective, but the threshold
    $0.3$ was selected before these ensembles were trained yet has not been
    validated on a dedicated held-out set; one partial check exists, in that the
    threshold predates the ten additional networks at $\Rey=300$, on which the
    accepted subset gives $75.9 \pm 6.3\%$ ($n=4$).}
    \label{tab:correction}
    \begin{tabular}{llccccccccc}
        \toprule
        & & & \multicolumn{4}{c}{single run} & \multicolumn{2}{c}{ensemble, all}
        & \multicolumn{2}{c}{ensemble, $R^2>0.3$} \\
        \cmidrule(lr){4-7}\cmidrule(lr){8-9}\cmidrule(lr){10-11}
        Geometry & $\Rey$ & $n_z$ & $R^2$ & $\omega_d\,[10^{-3}]$ & sinusoid & phase
        & $n$ & phase (all) & $n_{\mathrm{acc}}$ & phase ($R^2{>}0.3$) \\
        \midrule
        Circular & 300 & 4 & 0.59 & $-0.96$ & $+57.6$ & $+78.1$ & 20 & $51 \pm 25$ & \phantom{0}6 & $74 \pm 8$ \\
        Circular & 400 & 4 & 0.53 & $-1.03$ & $+55.4$ & $+82.8$ & 10 & $71 \pm 16$ & \phantom{0}8 & $78 \pm 6$ \\
        Circular & 800 & 8 & 0.33 & $+0.65$ & $+47.6$ & $+71.6$ & 10 & $41 \pm 42$ & \phantom{0}5 & $75 \pm 2$ \\
        Square   & 100 & 4 & 0.49 & $-0.55$ & $+54.0$ & $+82.5$ & 10 & $61 \pm 30$ & \phantom{0}6 & $80 \pm 4$ \\
        \bottomrule
    \end{tabular}
\end{table*}

\begin{table}[htbp]
    \centering
    \caption{Improvement as a function of prediction horizon, circular
    cylinder at $\Rey=400$. The single-run columns reproduce one representative
    training run; the final column is the mean\,$\pm$\,standard deviation over ten
    independently trained networks. Averaged over seeds the phase correction
    generally becomes more effective as the horizon lengthens ($61 \to 71\%$, with
    a slight dip at the final time), confirming that it does not degrade as the
    prediction is extended --- the behaviour expected of a correction that acts on a
    constant-rate drift rather than on a fitted envelope.}
    \label{tab:horizon}
    \begin{tabular}{lccc}
        \toprule
        & \multicolumn{2}{c}{single run} & 10-seed mean \\
        \cmidrule(lr){2-3}\cmidrule(lr){4-4}
        Steps beyond calibration & sinusoid & phase & phase \\
        \midrule
        \phantom{0}96 & $+59.9\%$ & $+74.8\%$ & $61 \pm 19$ \\
        192 & $+59.2\%$ & $+77.5\%$ & $66 \pm 19$ \\
        288 & $+59.5\%$ & $+78.9\%$ & $70 \pm 17$ \\
        385 & $+57.6\%$ & $+78.1\%$ & $71 \pm 16$ \\
        \bottomrule
    \end{tabular}
\end{table}

\subsection{The improvement survives reconstruction}
\label{sec:results:field}

The correction acts in the latent space, but the quantity of interest is the
reconstructed flow field. Table~\ref{tab:field} reports the field-space results.

Two figures are given, and the distinction between them matters. The relative
reduction in field RMSE against the uncorrected rollout is modest --
$8$--$15\%$. This is not a weakness of the correction but a consequence of the
autoencoder: most of the total field error is compression error, which no
temporal correction can address. The informative measure is $G$,
Eq.~\eqref{eq:gap-closed}: the fraction of the \emph{correctable} error -- the
excess of the uncorrected reconstruction over the compression floor -- that the
correction removes. By this measure the phase correction removes $91$--$98\%$ of
everything that is removable.

Two features of Table~\ref{tab:field} require comment. First, the ensemble
confirms the single-run values without inflating them: across the accepted
networks of this field-space ensemble $G = 91$--$98\%$, and the standard deviation
is one to two percentage points except at $\Rey = 300$. This ensemble is trained
separately from that of Table~\ref{tab:correction}, so its accepted counts
($3$, $5$, $5$, $10$) differ from the latent-space counts ($6$, $8$, $5$, $6$);
each is the $R^2>0.3$ subset of its own set of networks, and we do not treat them
as a single pooled population. We report ensemble improvements as mean\,$\pm$\,standard deviation for continuity
with the single-run values, but note that the percentage improvement is bounded
above by $100\%$ and unbounded below, so for the low-drift cases (notably the
all-seeds column at $\Rey=800$, $2\pm54$) the standard deviation is not a faithful
summary; the underlying error \emph{ratios}, which are strictly positive, are
better behaved, and we quote medians where the distribution is skewed. Second, $G$
is reported only on accepted networks, and this is a requirement rather than a
selection. The denominator of
Eq.~\eqref{eq:gap-closed} is the correctable error
$\varepsilon_{\mathrm{unc}} - \varepsilon_{\mathrm{CAE}}$, which vanishes when a
network happens to converge to a rollout that barely drifts; $G$ is then the ratio
of two vanishing quantities and is numerically meaningless. Unconditionally it
ranges from $-24\%$ to $+99\%$ at $\Rey = 300$, where three of ten networks have
drift ratios below $1.5$. On the square cylinder, where every network developed a
coherent drift and all ten are accepted (in the field-space ensemble), the
unconditional and conditional values coincide at $97.2 \pm 1.1\%$. The quantity is well posed precisely where the
diagnostic says the correction applies, which is the same condition under which it
is worth quoting.

In direct terms the phase correction reduces the field RMSE by $8$--$15\%$; the
sinusoid baseline reduces it by $8$--$14\%$ over the same cases. The two are close
in the field even though the phase correction's latent-space margin over the
sinusoid is $20.5$--$28.5$ points, because both remove the correctable temporal error
and are then bounded below by the same compression floor. The phase correction's
advantage is therefore genuine but is expressed in the latent space and in its
horizon behaviour (Table~\ref{tab:horizon}), not in the headline field-RMSE
figure; we state both so that neither is mistaken for the other.

The ratio $\rho$ confirms this. Before correction, $\rho = 0.43$--$0.64$
(ensemble means; $0.47$--$0.65$ in the single runs): a
substantial part of the field error is attributable to the temporal model. After
correction, $\rho = 0.06$--$0.13$: the residual field error is now almost
entirely compression error. The temporal model has been corrected to the limit of
what the representation permits.

\begin{table*}[htbp]
    \centering
    \caption{Field-space results over the extrapolation region, in per cent.
    \emph{phase} is the direct reduction in field RMSE relative to the uncorrected
    rollout; $G$ is the fraction of the \emph{correctable} field error removed,
    Eq.~\eqref{eq:gap-closed}. The ensemble columns are
    mean\,$\pm$\,standard deviation over the $n$ networks (of ten trained per case)
    that the $R^2$ criterion accepts. This field-space ensemble is trained
    independently of the latent-space ensemble of Table~\ref{tab:correction}: the
    networks share seed indices but not weights, so the two tables' accepted
    counts need not and do not coincide (here $3$, $5$, $5$, $10$ against $6$,
    $8$, $5$, $6$), each being the accepted subset of its own ensemble. Across the
    accepted networks the correction removes $91$--$98\%$ of the correctable
    error, driving $\rho$ from $0.43$--$0.64$ before correction to $0.06$--$0.13$
    after. $G$ is reported only on accepted networks because it is ill-conditioned
    otherwise (see text).}
    \label{tab:field}
    \begin{tabular}{llcccccc}
        \toprule
        & & & \multicolumn{2}{c}{single run} & \multicolumn{3}{c}{seed ensemble, $R^2>0.3$} \\
        \cmidrule(lr){4-5}\cmidrule(lr){6-8}
        Geometry & $\Rey$ & $n_z$ & phase & $G$ & $n$ & phase & $G$ \\
        \midrule
        Circular & 300 & 4 & $11.3$ & $95.4$ & \phantom{0}3 & $15.2 \pm 13.5$ & $91.1 \pm 7.5$ \\
        Circular & 400 & 4 & $15.5$ & $96.7$ & \phantom{0}5 & $\phantom{0}9.4 \pm 3.8$ & $97.5 \pm 1.0$ \\
        Circular & 800 & 8 & $\phantom{0}8.6$ & $92.0$ & \phantom{0}5 & $\phantom{0}9.7 \pm 3.6$ & $97.6 \pm 2.2$ \\
        Square   & 100 & 4 & $11.4$ & $98.8$ & 10 & $\phantom{0}8.2 \pm 6.1$ & $97.2 \pm 1.1$ \\
        \bottomrule
    \end{tabular}
\end{table*}

\begin{figure}[htbp]
    \centering
    \includegraphics[width=\linewidth]{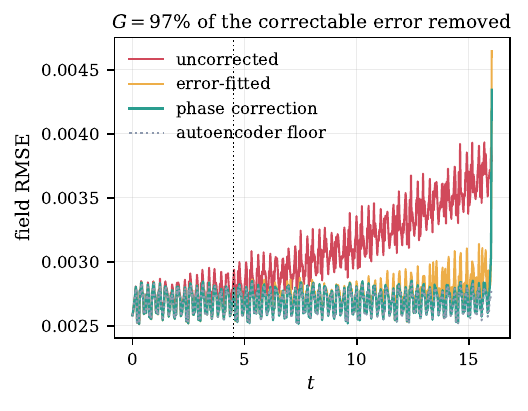}
    \caption{Field-space RMSE against the solver snapshots, circular cylinder,
    $\Rey=400$. The phase-corrected reconstruction lies essentially on the
    autoencoder floor -- the best any latent-space model could achieve.}
    \label{fig:field}
\end{figure}

\subsection{When the method applies, and how it knows}
\label{sec:results:applicability}

The correction is not universally applicable, and -- unlike the baseline -- it
signals its own inapplicability.

\paragraph{The phase-drift model must hold.}
The correction rests on $\phi(t)$ being linear with a resolvable slope. Where
$\omega$ is negligible, the regression that estimates it is fitting noise, and
the resulting correction injects phase error rather than removing it. This is
detected directly by the $R^2$ of that regression. Pooling the $50$ independently
trained networks of Table~\ref{tab:correction} gives a direct test of the
diagnostic, and the relationship is strong and monotone
(Table~\ref{tab:r2bands}): the Pearson correlation between calibration $R^2$ and
the achieved improvement is $r = 0.85$ ($p \approx 10^{-14}$; Spearman
$\rho = 0.93$). Because the fifty networks span two geometries, three Reynolds
numbers and two latent dimensions, a pooled correlation could in principle be
inflated by between-group variation. It is not: computed within each case
separately the coefficients are $r = 0.76$, $0.94$, $0.92$ and $0.96$ (all
$p < 10^{-3}$), and the group-centred correlation, which removes all
between-case variation, is $r = 0.85$ --- indistinguishable from the pooled
value. The relation holds within configurations, not merely across them. The efficacy rises from $19\%$ for $R^2 < 0.1$ to $77\%$ for
$R^2 > 0.3$, and---equally important---the scatter contracts monotonically over
the same range, from $\pm28$ to $\pm6$ percentage points. A high $R^2$ therefore
predicts not merely that the correction will work but that it will work
\emph{reliably}. The transition is a smooth ramp rather than a threshold effect;
the value $0.3$ is an operating point on that ramp, not a discontinuity. Because
$R^2$ grows with the calibration length as well as with the drift
(Table~\ref{tab:calib}), the numerical threshold is specific to the present
sampling and $N_{\mathrm{cal}}=150$; the transferable quantity is not $0.3$ itself
but the signal-to-noise interpretation behind it --- the drift must be resolvable
above the jitter of the phase estimate --- which a dimensionless slope $t$-statistic
would express directly. We use $R^2$ here for its transparency and flag the
non-dimensional form as the appropriate generalisation. The diagnostic requires no ground truth beyond the calibration
window already used to fit the correction, and it is available \emph{before} the
correction is applied---so conditioning on it is a prospective decision rule, not
a retrospective selection of favourable runs. One caveat on interpretation: $R^2$
is large precisely when the drift is large, and the reported efficacy is a
\emph{relative} error reduction, so accepted networks are also those with the most
correctable error. The gate should therefore be read as identifying where the
correction is worthwhile in relative terms, not as a claim that accepted networks
reach a lower absolute error than rejected ones; the absolute latent errors are
comparable across the two groups.

\begin{table}[htbp]
    \centering
    \caption{Correction efficacy as a function of the calibration-window $R^2$,
    pooled over all $50$ independently trained networks of
    Table~\ref{tab:correction} (both geometries, three Reynolds numbers, two
    latent dimensions). Efficacy increases and its scatter contracts monotonically
    with $R^2$.}
    \label{tab:r2bands}
    \begin{tabular}{lcc}
        \toprule
        $R^2$ band & $n$ & improvement (\%) \\
        \midrule
        $[0.0,\,0.1)$ & 14 & $19 \pm 28$ \\
        $[0.1,\,0.2)$ & \phantom{0}6 & $49 \pm 20$ \\
        $[0.2,\,0.3)$ & \phantom{0}5 & $52 \pm 13$ \\
        $\ge 0.3$     & 25 & $77 \pm 6$ \\
        \bottomrule
    \end{tabular}
\end{table}

\paragraph{Reynolds number versus latent dimension.}
The single $n_z=8$ case ($\Rey=800$) is also the weakest in
Table~\ref{tab:correction}, which alone cannot separate a Reynolds-number effect
from a latent-dimension effect. Holding $\Rey=800$ fixed and varying only the
latent dimension resolves it: at $n_z=4$ the improvement is $2.1\%$, at $n_z=8$ it
is $47.6\%$. The poor performance at low latent dimension is thus a compression
effect -- the four-dimensional latent space cannot represent this flow to the
accuracy the correction needs -- and not a property of the Reynolds number, a
distinction the ratio $\rho$ makes precise.

The same ratio identifies an upper boundary. Increasing the latent dimension to
$n_z = 16$ at $\Rey = 400$ removes the compression limitation entirely: the
rollout sits essentially on the reconstruction floor ($\rho = 0.09$), and with it
the phase structure disappears. No latent coordinate carries a coherent linear
drift ($R^2 = 0.01$ averaged over the sixteen coordinates, none exceeding $0.03$),
the extra coordinates holding low-amplitude incoherent content rather than a clean
oscillation. The diagnostic correctly reports this: with every $R^2$ far below the
threshold, the correction is not applied, and indeed there is almost nothing to
correct ($I_{\mathrm{pha}} = +6\%$). The phase-drift mechanism, and the correction
that exploits it, are thus properties of the compression-limited regime --- the
regime in which the autoencoder is doing non-trivial work and the dynamics are
forced onto a low-dimensional orbit. Below it (too few coordinates) the model is
compression-limited and there is no clean phase to correct; above it (too many)
the model is at its reconstruction floor and there is no drift to correct. The
correction is useful in the intermediate regime that a well-chosen ROM occupies,
and $\rho$ locates that regime without reference to ground truth.

The sinusoid baseline as implemented has no such check. This is a property of the
construction we compare against rather than of error-fitting in general: a
goodness-of-fit statistic for Eq.~\eqref{eq:sinusoid} could be defined and used
as a gate in the same way. What the comparison establishes is that the phase
correction's diagnostic comes for free --- the same regression that produces the
correction produces the criterion --- whereas for a fitted correction it would
have to be added deliberately. At $\Rey = 100$, $n_z = 8$, where the
network has learned the latent dynamics almost perfectly and there is
correspondingly little to correct, the sinusoid degrades the prediction by
$233\%$, while the phase correction -- whose fitted $\omega$ is near zero, as its
low $R^2$ correctly reports -- leaves it essentially unchanged.

\paragraph{Calibration must be sufficient.}
Table~\ref{tab:calib} shows the dependence on the length of the calibration
window. Counter-intuitively, the single-parameter method requires \emph{more}
calibration data than the three-parameter baseline, not less: $\omega$ is a rate,
estimated as the slope of a phase difference that accumulates to only
$\sim 10^{-2}$\,rad over the entire rollout, and short windows cannot resolve it
above the jitter of the instantaneous-phase estimate. Below $\approx 75$ steps
($\approx 4$ shedding cycles) the phase correction is worse than the sinusoid and
can be severely harmful; beyond $\approx 150$ steps both methods saturate.

This provides a test of the $R^2$ diagnostic that is independent of the seed
ensemble, since here the network is fixed and only the calibration budget varies.
The diagnostic passes: the catastrophic $-257\%$ degradation at
$N_{\mathrm{cal}}=25$ is accompanied by $R^2 = 0.167$, below the threshold, and
every window the criterion accepts yields at least $+76\%$. Within this sweep
there are no false acceptances, though six calibration lengths at a single
Reynolds number are too few to establish that as a general property. The criterion is conservative rather than sharp --- at
$N_{\mathrm{cal}}=75$ it rejects a window that would in fact have helped
($+58.9\%$) --- which is the appropriate direction of error for a gate whose
purpose is to prevent harm.

\paragraph{The dynamics must be periodic.}
\label{sec:results:breakdown}
At $\Rey = 1000$ the wake is temporally broadband and the instantaneous phase is
not well defined. The decomposition of Sec.~\ref{sec:methods:decomp} returns
variance shares exceeding $100\%$, indicating that the analytic-signal
representation on which it rests has broken down, and the fitted $\omega$ is
larger than at any other Reynolds number by three orders of magnitude. We
therefore exclude $\Rey = 1000$ from the phase analysis. The correction is
defined for flows possessing a well-defined dominant frequency, and this case
lies outside that class. We note the limits of this observation: a
two-dimensional wake at $\Rey = 1000$ is not a physically realisable flow, so
while it demonstrates that the diagnostic identifies the breakdown of its own
representation, it does not establish how the method behaves for genuinely
broadband or quasi-periodic dynamics. That test requires a flow whose loss of
periodicity is physical rather than numerical, and we have not performed it.

\begin{table}[htbp]
    \centering
    \caption{Dependence on calibration length (circular cylinder, $\Rey=400$,
    $n_z=4$). The phase correction requires a longer window than the baseline
    because it estimates a rate rather than a magnitude. The $R^2$ column shows
    that the diagnostic detects this unaided: every window it rejects
    ($R^2<0.3$) is one on which the correction underperforms the baseline or is
    harmful, and every window it accepts gives $+76\%$ or better.}
    \label{tab:calib}
    \begin{tabular}{cccccc}
        \toprule
        $N_{\mathrm{cal}}$ (steps) & cycles & $R^2$ & sinusoid & phase & verdict \\
        \midrule
        \phantom{0}25 & 1.4 & 0.167 & $+25.9$ & $-256.6$ & reject \\
        \phantom{0}50 & 2.7 & 0.211 & $+37.0$ & $\phantom{-00}-9.4$ & reject \\
        \phantom{0}75 & 4.1 & 0.246 & $+45.2$ & $\phantom{-0}+58.9$ & reject \\
        100 & 5.5 & 0.304 & $+50.2$ & $+76.4$ & accept \\
        150 & 8.2 & 0.530 & $+55.4$ & $+82.8$ & accept \\
        200 & 11.0 & 0.681 & $+54.9$ & $+84.2$ & accept \\
        \bottomrule
    \end{tabular}
\end{table}

\subsection{A linear propagator does not suffice}
\label{sec:results:dmd}

\begin{figure*}[htbp]
    \centering
    \includegraphics[width=\textwidth]{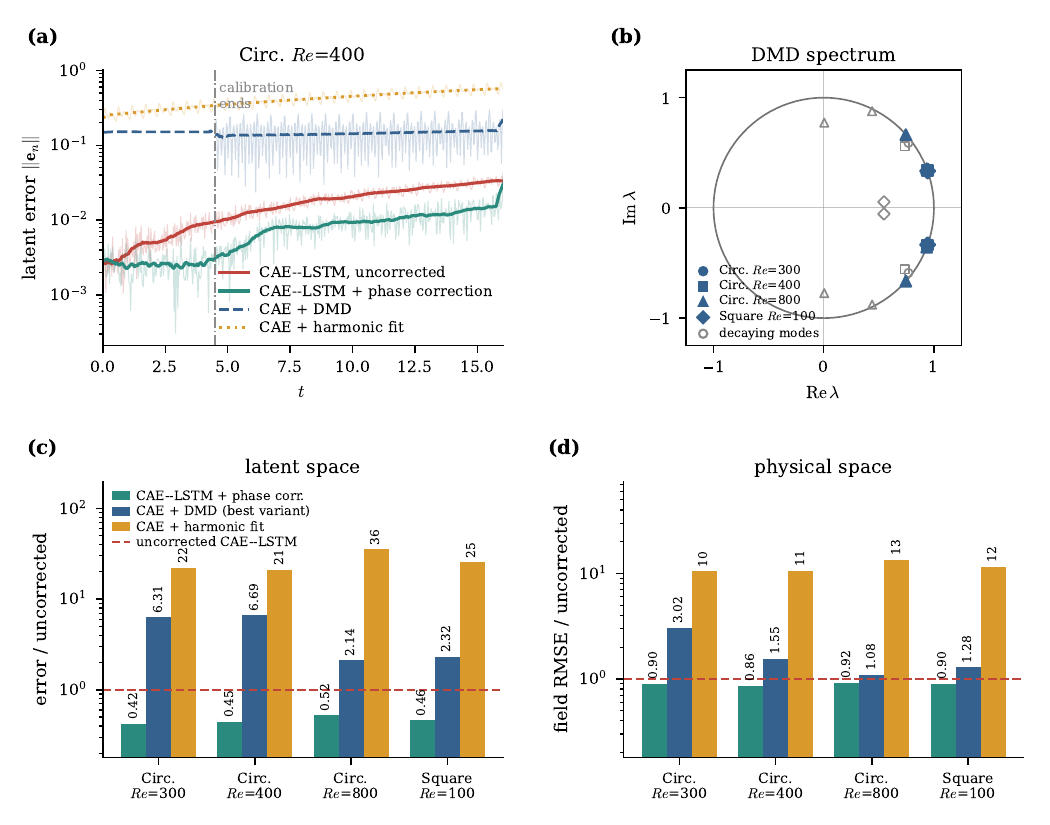}
    \caption{Linear-propagator baselines. (a) Latent error against time for the
    circular cylinder at $\Rey=400$; faint traces are the raw error, bold traces a
    moving average over one shedding period. (b) DMD eigenvalues in the complex
    plane; filled markers denote the neutrally stable pair
    ($|\lambda| > 0.995$) carrying the shedding frequency, open markers the damped
    harmonics. (c),~(d) Error over the extrapolation region relative to the
    uncorrected {CAE--LSTM}, in the latent space and in physical space; DMD is
    shown at the best of its three variants.}
    \label{fig:dmd}
\end{figure*}

The narrowband character of the latent signals invites the objection that a
linear propagator would do as well or better, and without any phase drift to
correct. We test this directly, replacing only the temporal model and retaining
the same autoencoder (Sec.~\ref{sec:methods:dmd}). The comparison is summarised in
Table~\ref{tab:dmd} and Fig.~\ref{fig:dmd}.

The linear operator is well conditioned and correctly fitted. In every case its
spectrum contains a conjugate pair on the unit circle
($|\lambda| \approx 1$) whose frequency reproduces the shedding Strouhal number of
the flow to within the precision the grid supports
(Fig.~\ref{fig:dmd}\,b): $\St_{\mathrm{DMD}} \approx 0.21$, $0.22$, $0.23$ and
$0.15$, matching the measured values. That the linear operator recovers the
correct frequency confirms it is properly fitted; its failure is one of
propagation, not identification. The
remaining eigenvalues are strongly damped and correspond to harmonics. DMD
identifies the physics; what it cannot do is propagate it.

Over the extrapolation region the best of the three DMD variants is
$2.1$--$6.7$ times less accurate than the uncorrected network and
$7.5$--$39$ times less accurate than the corrected one. The harmonic model is
worse still, by a factor of $21$--$36$. The margin is not an artefact of the
latent norm: in physical space the linear baselines are also worse than the
uncorrected network in every case.

The three DMD variants behave similarly, which is itself diagnostic. Projecting
the eigenvalues onto the unit circle removes all growth and decay, yet does not
improve matters --- it is marginally better at $\Rey=300$ and $400$, marginally
worse at $\Rey=800$, and substantially worse for the square cylinder, where
forcing a marginal mode onto the unit circle destabilises the reconstruction. The
failure is therefore not simply the decay of damped modes. It is that a single
linear operator cannot represent the latent dynamics to the accuracy the recurrent
model attains: fitted on the training window, the operator leaves a one-step
residual of $2.9\times10^{-2}$, some thirty times the network's. Refitting to
include the calibration window (the variant given the same ground truth the phase
correction uses) helps in three cases of four but never by more than a factor of
$1.5$.

The reason is visible in the residuals. The harmonic model of
Eq.~\eqref{eq:harmonic} leaves a root-mean-square residual of $0.068$--$0.099$ on
the data it was \emph{fitted} to, whereas the network's error over the
extrapolation region it has never seen is $0.018$--$0.024$: the fixed-frequency
model cannot match, on data it has seen, what the network achieves on data it has
not. This is quantitatively consistent with the in-band energy fraction of
Sec.~\ref{sec:methods:decomp}: a residual of $0.099$ against unit signal
root-mean-square corresponds to $99.0\%$ in-band energy, and one of $0.068$ to
$99.5\%$. The latent dynamics are narrowband to one part in a hundred, which is
ample for the analytic-signal representation, and insufficient by an order of
magnitude for a fixed-frequency propagator to compete.

Note that the in-fit residual and the extrapolation error are not directly
comparable quantities: the harmonic model's error over the extrapolation region is
far larger than its in-fit residual, because it accumulates its own phase drift
(below). The comparison above is deliberately conservative --- it sets the
harmonic model's best case against the network's worst.

A second observation bears directly on the generality of the mechanism. The
harmonic model, whose amplitude is fixed by construction so that phase error is
its only available failure mode, exhibits exactly the linear phase drift
described in Sec.~\ref{sec:results:linear}, at a rate $1.7$--$19.8$ times larger
than the network's and with comparable or higher $R^2$ ($0.21$--$0.62$). DMD,
by contrast, shows negligible coherent phase drift ($R^2 \le 0.20$) because its
error is dominated by the decay of its damped modes --- an amplitude failure, and
one that the decomposition of Sec.~\ref{sec:results:decomp} correctly attributes
to amplitude rather than phase. The drift is therefore not a peculiarity of
recurrent architectures: it appears in a non-recurrent, non-learned propagator
whenever the amplitude is held correct and the frequency is estimated from data.

\begin{table*}[htbp]
    \centering
    \caption{Linear-propagator baselines sharing the same autoencoder, scored in the
    latent space over the extrapolation region and expressed as a ratio to the
    uncorrected {CAE--LSTM} (values below unity are better than the uncorrected
    network; the metric is that of Eq.~\eqref{eq:improvement} throughout). All
    three DMD variants are shown separately: the operator fitted on the training
    window, the same operator with its eigenvalues projected onto the unit circle,
    and the operator refitted to include the calibration window.
    $\St_{\mathrm{DMD}}$ is the frequency of the DMD eigenvalue pair nearest the
    unit circle, compared with the measured shedding Strouhal number $\St$; both
    are quoted to the precision the grid supports
    (Sec.~\ref{sec:methods:convergence}).}
    \label{tab:dmd}
    \begin{tabular}{llccccccc}
        \toprule
        & & \multicolumn{2}{c}{{CAE--LSTM}} & \multicolumn{3}{c}{DMD variant}
        & & \\
        \cmidrule(lr){3-4}\cmidrule(lr){5-7}
        Geometry & $\Rey$ & sinusoid & phase & fitted & unit circle & $+$calib.
        & harmonic & $\St_{\mathrm{DMD}}$ / $\St$ \\
        \midrule
        Circular & 300 & 0.42 & 0.219 & 9.16 & 9.67 & 6.31 & 22.3 & 0.213 / 0.213 \\
        Circular & 400 & 0.45 & 0.172 & 7.85 & 7.76 & 6.69 & 21.2 & 0.219 / 0.219 \\
        Circular & 800 & 0.52 & 0.284 & 2.74 & 2.53 & 2.14 & 35.7 & 0.232 / 0.232 \\
        Square   & 100 & 0.46 & 0.175 & 2.32 & 10.0\phantom{0} & 2.60 & 25.4 & 0.149 / 0.148 \\
        \bottomrule
    \end{tabular}
\end{table*}

\subsection{Periodic extension, and the limits of a stationary test}
\label{sec:results:periodic}

A linear propagator is not the only model-free predictor available for a periodic
flow. The simplest of all is to tile the last observed period of the trajectory
forward. On a saturated limit cycle this needs no model, no latent space and no
training --- only an estimate of the period from the calibration window --- and it
is the baseline a reader will reach for the moment the latent dynamics are
described as nearly single-frequency. We evaluate it in both spaces, estimating
the period from the calibration window by an interpolated periodogram peak. The
period must be resolved to sub-sample accuracy: rounding it to the nearest sample
injects an artificial phase error of the very kind this paper concerns, and
inflates the tiling error by more than an order of magnitude. With a fractional
period the comparison is fair.

In the latent space the phase-corrected model wins decisively. Periodic tiling
gives error ratios of $0.96$--$1.41$ relative to the uncorrected network --- that
is, it is comparable to making no prediction beyond the limit cycle at all, and in
three of four cases worse than the uncorrected rollout --- against $0.17$--$0.28$
for the phase correction, a factor of $5$--$6$. The tiling is worst at
$\Rey = 800$ ($1.41$), the least cleanly periodic case, which is the expected
behaviour of a predictor that assumes exact periodicity.

In physical space the result is closer and, for one case, unfavourable, and we
report it in full because it bears directly on how the contribution should be
read. Table~\ref{tab:periodic} gives the field-RMSE reduction of model-free field
tiling against the phase-corrected model. For the three circular cylinders the
phase correction wins ($+11.5$ vs $+8.8\%$, $+15.7$ vs $+12.5\%$, $+8.7$ vs
$-2.1\%$), and at $\Rey = 800$ tiling fails outright ($-2.1\%$) because that flow
is the least periodic. For the square cylinder, however, tiling the raw solver
snapshots forward ($+13.6\%$) outperforms the entire pipeline ($+11.6\%$) --- and
indeed exceeds the CAE reconstruction floor ($+11.7\%$), because a model-free field
prediction never passes through the autoencoder and is therefore not bounded by
its compression error.

\begin{table}[htbp]
    \centering
    \caption{Periodic extension versus the phase-corrected model, as field-RMSE
    reduction over the extrapolation region (per cent, relative to the uncorrected
    rollout). \emph{Tiling} repeats the last observed period of the solver
    snapshots directly, with no model. The phase correction wins on the three
    circular cylinders; on the square cylinder, the most cleanly periodic flow,
    model-free tiling wins and exceeds even the autoencoder's reconstruction
    floor.}
    \label{tab:periodic}
    \begin{tabular}{llccc}
        \toprule
        Geometry & $\Rey$ & tiling (field) & phase corr. & CAE floor \\
        \midrule
        Circular & 300 & $+8.8$  & $+11.5$ & $+12.0$ \\
        Circular & 400 & $+12.5$ & $+15.7$ & $+16.3$ \\
        Circular & 800 & $-2.1$  & $+8.7$  & $+9.4$ \\
        Square   & 100 & $+13.6$ & $+11.6$ & $+11.7$ \\
        \bottomrule
    \end{tabular}
\end{table}

This comparison is deliberate. A single stationary limit cycle is the regime most
hostile to a learned reduced-order model and most favourable to trivial tiling:
there is no parameter to vary, no transient to capture and no three-dimensionality,
so the one thing the model offers --- a compressed dynamical representation that
generalises beyond a single periodic orbit --- is precisely the thing a stationary
test cannot reward, while tiling need only assume the flow repeats, which here it
does. That the correction \emph{matches} tiling even in this most unfavourable
setting, and then (Sec.~\ref{sec:results:modulated}) continues to work when tiling
collapses the moment periodicity is relaxed, is the point. Periodic extension is
close to the best a predictor can do on an exactly-repeating flow; its usefulness
ends exactly where the flow stops repeating. We therefore do not claim that the pipeline
predicts these particular \emph{stationary} wakes more accurately than a periodic
baseline; on the square cylinder it does not. The distinction the phase-drift
analysis draws is not between our correction and tiling on a single limit cycle ---
where both succeed --- but between flows that repeat exactly and flows that do not.
Tiling requires exact periodicity; the phase correction requires only coherence.
The $\Rey = 800$ result, where tiling already degrades on a merely
\emph{less}-periodic wake, indicates how quickly that requirement bites. The next
section makes the separation explicit, on a flow constructed so that it never
repeats.

\subsection{A non-stationary wake: where periodic extension fails and the
correction does not}
\label{sec:results:modulated}

The comparison above leaves one question open. On a stationary limit cycle the
phase correction and periodic tiling both work, and the argument that the
correction is nonetheless the more general tool rests on the claim that tiling
requires exact periodicity while the correction requires only coherence. That claim
can be tested directly, by constructing a flow that is coherent but not periodic.

We drive the $\Rey = 400$ circular cylinder with a slowly varying inflow,
$U(t) = U_0\,[\,1 + \varepsilon \sin(2\pi f_m t)\,]$ with $\varepsilon = 0.05$ and a
modulation period $1/f_m \approx 15$ shedding cycles, so that the instantaneous
Reynolds number drifts gently between $380$ and $420$. The shedding remains a clean,
coherent oscillation at every instant --- the instantaneous amplitude varies by only
$\pm 12\%$ about its mean --- but its frequency now tracks the inflow, so the
shedding period changes continuously and no cycle reproduces the next. This is
precisely the regime the previous section identified as the one that separates the
two predictors: coherent enough that an instantaneous phase is well defined, not
periodic enough that a fixed template can be tiled. A reduced-order model is trained
on this flow exactly as for the stationary cases, with no modification, and rolled
out autoregressively.

The outcome is unambiguous (Table~\ref{tab:modulated}). Periodic extension, which
must assume a single representative period, fails by more than an order of
magnitude: tiling the last observed period forward gives a latent error $31$ times
larger than making no correction at all, and a field error over twenty times the
uncorrected value, because the period it copies is already stale by the time it is
applied. The phase correction, by contrast, remains effective. It removes $49\%$ of
the latent rollout error and $42\%$ of the field error --- less than the
$74$--$80\%$ it achieves on the stationary cases, as expected, since the drift is
no longer perfectly linear and a single fitted rate cannot track a continuously
varying frequency exactly, but far from the failure of the model-free baseline. The
correction works because what it acts on --- the instantaneous phase of a coherent
oscillation --- still exists when exact periodicity does not.

\begin{table}[htbp]
    \centering
    \caption{The non-stationary (modulated-inflow) wake. Error over the
    extrapolation region relative to the uncorrected rollout: latent as a ratio
    (below one is better), field as a percentage reduction (positive is better).
    Periodic tiling assumes a single period and fails once the period drifts; the
    phase correction, requiring only a coherent instantaneous phase, remains
    effective.}
    \label{tab:modulated}
    \begin{tabular}{lcc}
        \toprule
        & latent (ratio) & field (\% reduction) \\
        \midrule
        uncorrected              & $1.00$  & $0$ \\
        periodic tiling          & $31.1$  & $-2254$ \\
        phase correction         & $0.51$  & $+42.3$ \\
        \bottomrule
    \end{tabular}
\end{table}

Two points should be made honestly. First, the correction is less effective here
than on the stationary flows, and it must be: a drift rate fitted on the
calibration window is a constant, and the true drift is now itself time-varying, so
a residual remains that a constant realignment cannot remove. A correction that
tracked the instantaneous frequency would recover more, and we note it as the
natural extension (Sec.~\ref{sec:discussion:limits}); the constant-rate correction
is reported here because it is the method of the rest of the paper, applied
unchanged. Second, this is a single constructed case, offered as an existence proof
that the coherence-versus-periodicity distinction is real and operational, not as a
survey of non-stationary flows. What it establishes is that the phase-drift picture
is not an artefact of the stationary limit cycles on which it was found: the error
remains a phase error, and remains correctable as one, when the flow itself refuses
to repeat.

\subsection{Drift magnitude varies across training runs; its structure does not}
\label{sec:results:seeds}

How much phase drift accumulates during a rollout is not a fixed property of the
flow. Retraining the LSTM on identical data with different random
initialisations, holding the autoencoder fixed, produces rollouts whose drift
varies by nearly an order of magnitude: across forty seeds at $\Rey = 500$ the
ratio of the error at the end of the rollout to that at its start has mean $5.7$
and standard deviation $4.0$ (coefficient of variation $0.70$), despite all forty
networks reaching essentially identical one-step validation error
($\sim\!10^{-6}$).

We searched for a property of the trained network that would predict its drift,
testing the autoencoder reconstruction error, the LSTM one-step validation error,
the network weight norm, the ratio of predicted to true limit-cycle amplitude, and
the predicted shedding frequency (Table~\ref{tab:predictors}). None correlates with
the drift: every Pearson coefficient satisfies $|r| < 0.15$, none approaches a
Bonferroni-corrected significance threshold, and the Spearman rank correlations and
the phase-drift rate give the same result. This is not a low-power null. With forty
seeds the analysis resolves correlations down to $|r| = 0.31$ ($0.40$ after
Bonferroni correction for the five predictors), so the confidence intervals bound
the size of any undetected effect rather than reflecting an inability to detect
one. We state the bound conservatively: $|r| = 0.31$ is the smallest correlation
that would \emph{reach} significance at $n = 40$, whereas reliable detection
(power $\ge 0.8$) requires a true correlation nearer $|r| \approx 0.43$. The
appropriate conclusion is therefore that none of the five predictors tested has
$|r| \gtrsim 0.4$; neither arbitrarily weak dependence nor a predictor outside
this set has been excluded. The strongest
candidate in a preliminary ten-seed analysis---the limit-cycle
amplitude ratio, at $r = -0.43$---regresses to $r = 0.13$ at forty seeds,
consistent with the earlier value being a sampling artefact.

The direction of the drift is likewise not fixed. Of the four cases in
Table~\ref{tab:correction} the network's clock runs slow in three
($\omega_d < 0$) and fast in one ($\Rey = 800$, $\omega_d = +6.5\times10^{-4}$),
so the sign is a property of the particular trained model rather than a systematic
bias of the architecture.

What does \emph{not} vary is the character of the error. In every seed the error
spectrum peaked at the shedding frequency and the decomposition returned a phase
share above $95\%$. The \emph{structure} of the error is a reproducible property
of the flow; its \emph{magnitude} is a property of the particular training run.

The consequence for the method is a conditional statement, and we make it
explicitly: the correction removes the phase drift that a given rollout has
accumulated. Where a network has drifted, it removes most of it; where a network
has happened to converge to a stable rollout, there is little to remove, and
$R^2$ reports this before any correction is attempted. The drift magnitude need
not be predictable for the method to work, because it is measured directly from
the calibration window of each rollout rather than forecast in advance.

\begin{table}[htbp]
    \centering
    \caption{Correlation between candidate predictors and observed rollout drift
    across forty random initialisations ($\Rey=500$). No predictor reaches
    significance; at this sample size correlations are resolvable down to
    $|r|=0.31$ and reliably detectable at $|r|\approx0.43$, so these values bound
    any moderate undetected effect rather than reflecting low power. Entries are
    Pearson coefficients;
    Spearman rank correlations agree.}
    \label{tab:predictors}
    \begin{tabular}{lcc}
        \toprule
        Predictor & Pearson $r$ & $p$ \\
        \midrule
        CAE reconstruction error        & $+0.02$ & $0.92$ \\
        One-step validation error       & $-0.06$ & $0.70$ \\
        LSTM weight norm                & $+0.03$ & $0.86$ \\
        Limit-cycle amplitude ratio     & $+0.13$ & $0.44$ \\
        Predicted shedding frequency    & $-0.15$ & $0.36$ \\
        \bottomrule
    \end{tabular}
\end{table}
\section{Discussion}
\label{sec:discussion}

\subsection{Why the error appears at the shedding frequency}
\label{sec:discussion:mechanism}

The spectral peak of Sec.~\ref{sec:results:spectrum} is not a property of the
architecture. Given the decomposition of Sec.~\ref{sec:results:decomp} -- that
the error is a phase error and nothing else -- it follows algebraically.

Let a latent coordinate evolve on the limit cycle as
\begin{equation}
    z(t) = A \cos\!\left( 2\pi f_s t \right),
    \label{eq:true-osc}
\end{equation}
and let the network reproduce this waveform, at the correct amplitude, but with
an accumulated phase error $\phi(t)$:
\begin{equation}
    \hat{z}(t) = A \cos\!\left( 2\pi f_s t + \phi(t) \right).
    \label{eq:pred-osc}
\end{equation}
The measured amplitude fidelity --- a root-mean-square deviation of $0.15\%$
(Table~\ref{tab:decomposition}) --- is what licenses writing the same $A$ in both.
The rollout error is then, by the sum-to-product identity,
\begin{equation}
    e(t) = \hat{z}(t) - z(t)
    = -2A \sin\!\left( \frac{\phi(t)}{2} \right)
          \sin\!\left( 2\pi f_s t + \frac{\phi(t)}{2} \right).
    \label{eq:error-decomp}
\end{equation}
Equation~\eqref{eq:error-decomp} accounts for both observations of
Sec.~\ref{sec:results}. The second factor oscillates at the shedding frequency:
the error must appear in the spectrum as a peak at $f_s$, whatever the value of
$\phi$. The first factor is a slowly varying envelope set by the accumulated
phase error: as the prediction slips further out of phase the error grows,
saturating only when prediction and truth are in antiphase.

We note that the spectral peak and the amplitude--phase partition are not
independent findings. Once the predicted amplitude is known to be correct to
within $0.15\%$, a phase-only error on a limit cycle is algebraically obliged to
peak at $f_s$; Eq.~\eqref{eq:error-decomp} exhibits precisely this. The spectral
observation does real work---it establishes that the error is a coherent
oscillation rather than broadband noise, which the independent-error picture would
not predict---but the informative and non-obvious result is the partition itself,
and it is on that, rather than on the location of the peak, that the correction
rests.

The measured phase drift is remarkably small -- some $0.02$--$0.03$\,rad, or
$0.003$--$0.005$ of a cycle, over an entire rollout. For $\phi \ll 1$,
Eq.~\eqref{eq:error-decomp} linearises to
\begin{equation}
    e(t) \approx -A\,\phi(t)\, \sin\!\left( 2\pi f_s t \right),
    \label{eq:error-linear}
\end{equation}
so that the error is a sinusoid at $f_s$ whose amplitude is directly proportional
to $\phi$. Since $\phi$ grows linearly (Sec.~\ref{sec:results:linear}), the error
grows linearly, and an $O(10^{-2})$ phase slip on an $O(1)$ oscillation produces
an $O(10^{-2})$ error -- which is precisely the order of the observed drift. The
smallness of the phase error and the largeness of its consequence are not in
tension: they are related by the amplitude of the signal on which the phase error
acts.

Two things follow, and both are borne out.

First, the mechanism of long-horizon drift on a periodic flow is not the
accumulation of independent errors but the accumulation of \emph{phase}. The
network has learned the attractor; it has not learned to traverse it at exactly
the right rate.

Second, Eq.~\eqref{eq:error-linear} explains why the sinusoid baseline works at
all. It is, in effect, the first-order expansion of the phase-drift error, fitted
empirically. That it is outperformed by $20.5$--$28.5$ percentage points
(Table~\ref{tab:correction}) by a method that acts on $\phi$ directly is then
unsurprising: the baseline models the consequence, and must extrapolate an
envelope it has fitted; the phase correction models the cause, and extrapolates a
constant. The horizon dependence of Table~\ref{tab:horizon} is the visible
signature of that difference.

\subsection{A caution regarding the detuning}
\label{sec:discussion:detuning}

A natural reading of Eq.~\eqref{eq:error-linear} is that $\phi(t) = 2\pi\,\delta
f\, t$, where $\delta f$ is the difference between the network's effective
oscillation frequency and the true shedding frequency, and hence that $\delta f$
should predict the drift. It does not, and the discrepancy is worth stating.

Estimating $\delta f$ as the difference of the median instantaneous frequencies
of prediction and truth yields values that do not reproduce the measured slope of
$\phi(t)$, and that do not correlate with observed drift across training seeds
(Table~\ref{tab:predictors}). The reason is one of measurement rather than
physics: $\delta f$ is a difference of two nearly equal quantities, each estimated
with jitter comparable to the difference itself, whereas the slope of $\phi(t)$ is
a direct regression over the whole trajectory and is far better conditioned. The
drift rate $\omega$ used throughout this work is therefore defined as that slope,
and not as a difference of frequency estimates. We note this explicitly because
the frequency-difference route is the more obvious one and, in our experience,
the more misleading.

\subsection{Relation to existing work}
\label{sec:discussion:prior}

Rollout instability in autoregressive surrogates is well
documented\cite{vlachas2018lstm} and a range of remedies exist. They fall broadly
into three groups.

\emph{Training-side remedies} modify how the model is fitted: multi-step or
``pushforward'' losses that unroll the model during training, noise injection to
make the network robust to its own errors, and spectral loss terms that penalise
mismatched frequency content. These are effective but require retraining and
alter the model itself\cite{brandstetter2022mppde, sanchez2020simulate, bengio2015scheduled}.
For the most common of them we can be concrete: injecting Gaussian noise into the
network's inputs during training does not remove the drift. We train a ten-seed
ensemble at each of four noise levels at $\Rey = 400$ (forty networks in all),
spanning from no injection up to the level at which the one-step validation error
begins to degrade. The mean drift ratio stays at $6$--$8$ and the
fitted rate $\omega_d$ is if anything larger, not smaller, so the phase
realignment remains both applicable and necessary. This does not exhaust the
training-side category --- pushforward and spectral losses may behave differently
--- but it rules out the specific remedy most likely to be proposed as a
substitute for the correction, on the flow where the correction works best.

\emph{Architectural remedies} constrain the latent dynamics -- Koopman
formulations that seek coordinates in which the dynamics are linear, or operator
methods that parameterise the evolution in the Fourier domain. These change what
is learned rather than correcting what was learned\cite{lusch2018koopman, li2021fourier}.

\emph{Post-hoc corrections} act on a frozen model, typically by training a second
network to regress the residual\cite{wang2020rnnclosure}. These are closest in spirit to the present work,
but they treat the error as an unstructured quantity, and consequently require a
model of complexity comparable to the one being corrected. We stress that the
sinusoid of Eq.~\eqref{eq:sinusoid} is a deliberately minimal reference -- the
correction the spectral observation alone would suggest -- and not a learned
closure of this kind; the comparison isolates the value of acting on the phase.
Its large latent-space margin should be read in that light, the more so as it
narrows to a few percentage points in field space (Sec.~\ref{sec:results:field}).

What distinguishes the present approach is what is done \emph{before} correcting.
One qualification on the two-geometry evidence should be made where the claim is
first advanced rather than only here. Section~\ref{sec:discussion:mechanism} shows that any
correct-amplitude, phase-drifting prediction of a limit cycle must produce an
error spectrum peaked at the fundamental. The square cylinder therefore confirms
an algebraic consequence, not an independent hypothesis, and the spectral result
should not be read as a generalisation test. What the second geometry does test,
and what genuinely requires multiple cases, is the amplitude--phase partition
itself and the linearity of $\phi(t)$: neither is guaranteed by the algebra, and
both hold across two shedding mechanisms and a factor of $1.5$ in Strouhal number.

It is worth distinguishing the present finding from spectral bias, the documented
tendency of neural networks to learn low-frequency content preferentially and to
under-resolve high frequencies\cite{rahaman2019spectral}. In autoregressive
rollouts that bias is normally reported as a progressive attenuation of
small-scale structure --- an \emph{amplitude} effect, producing predictions that
grow smooth. The mechanism described here is the opposite in character: the
amplitude of the dominant mode is preserved to within $0.15\%$, and the error is
concentrated in the \emph{phase} of that same mode. The two are complementary
failure modes rather than competing descriptions of one, and a surrogate could
exhibit either or both.

The realignment is also close in spirit to the method of slices, or
template-fitting, used to factor out a continuous symmetry in spatiotemporal
dynamics\cite{budanur2015}: there a phase is removed to expose the shape dynamics
on a limit cycle, here a phase \emph{error} is removed to expose that the shape
dynamics were correct all along. Amplitude--phase error decomposition is itself
long established: the partition of
numerical error into dissipative (amplitude) and dispersive (phase) parts is
classical in advection-scheme analysis\cite{takacs1985}, and phase alignment of
travelling and oscillatory structures underlies symmetry-reduction and
template-fitting methods in reduced-order modelling\cite{rowley2000symmetry}. What
is new is not the decomposition or the idea of phase alignment, but their
application to the autoregressive rollout error of a latent-space neural
reduced-order model: to our knowledge this error has not previously been
decomposed quantitatively in this way, the specific finding that it is
$97$--$98\%$ phase has not been reported, and the resulting one-parameter,
self-diagnosing correction has no direct precedent (we discuss the recent work that
reaches the qualitative observation below). Once the decomposition is established, the correction requires no
auxiliary network, no retraining, and a single parameter per latent coordinate.

That phase, rather than amplitude, governs the long-horizon degradation of these
models is consistent with a growing body of recent work, which we do not claim to
pre-date. Solera-Rico \textit{et al.}\cite{solerarico2026compactness} report that
convolutional-autoencoder latent coordinates for controlled wakes carry more
irregular, broadband dynamics than their proper-orthogonal-decomposition
counterparts, so that autoregressive forecasts degrade faster --- a
compression-versus-predictability trade-off whose mechanism they attribute to the
faster accumulation of phase errors. Viknesh and
Arzani\cite{viknesh2026diano}, studying autoregressive latent rollouts of
cylinder-wake flow, similarly find the dominant error to arise from progressive
phase shifts and positional drift of the shed vortices rather than from amplitude
or core-structure errors. Our contribution is not the qualitative observation that
phase dominates --- which these and other studies reach independently --- but its
quantification: the amplitude/phase variance decomposition that assigns
$97$--$98\%$ of the error to phase, the demonstration that the rollout-error
spectrum concentrates at the shedding frequency across two shedding mechanisms, the
algebraic account of why it must, and the one-parameter, self-diagnosing correction
that follows. Where the prior work establishes \emph{that} phase matters, we
establish \emph{how much}, \emph{why}, and \emph{what to do about it}.

These training-side results make the relationship concrete rather than assumed.
The relationship to spectral-loss methods is complementary rather than
competitive. A frequency-aware training loss would be expected to reduce the phase
drift rate $\omega$ at source; the present correction removes the consequences of
whatever drift remains. Combining the two is a natural direction and is not
pursued here.

\subsection{Limitations}
\label{sec:discussion:limits}

\paragraph{The flow must possess a well-defined phase.}
The decomposition and the correction both rest on the analytic-signal
representation, which requires a dominant frequency. At $\Rey = 1000$, where the
two-dimensional wake is temporally broadband, the decomposition returns variance
shares exceeding $100\%$ -- an unambiguous signal that the representation has
broken down -- and the method is not applicable. We make no claim regarding
turbulent flows.

\paragraph{The simulations are two-dimensional.}
Both configurations suppress spanwise instabilities present in the physical flow
above $\Rey \approx 190$ (circular) and $\Rey \approx 150$--$200$ (square). This is
deliberate: enforcing two-dimensionality preserves the clean periodicity in which
the mechanism is visible. It also means the higher-Reynolds-number cases here are
not physically realisable wakes. Whether the phase-drift mechanism survives genuine
three-dimensionality is open, and we regard it as the most important question this
work leaves unanswered.

\paragraph{The benefit is conditional on drift being present.}
As Sec.~\ref{sec:results:seeds} establishes, how much phase drift a given rollout
accumulates is not reproducible across training runs. The correction removes the
drift that is there; it does not guarantee a fixed improvement. We regard the
honest statement of the method's benefit as conditional, and we rely on $R^2$ to
establish whether the condition is met before the correction is applied.

\paragraph{The correction requires a calibration window with ground truth.}
As with any post-hoc correction, $\omega$ must be estimated somewhere. We use 150
steps ($\approx 8$ shedding cycles); Sec.~\ref{sec:results:applicability} shows
that fewer than $\approx 75$ is actively harmful. The method therefore reduces,
but does not eliminate, the dependence on ground truth: it converts a requirement
for truth over the whole horizon into a requirement for truth over a few cycles.

\paragraph{A constant drift rate cannot track a varying frequency.}
The correction estimates one drift rate per latent coordinate and applies it as a
constant. On a stationary limit cycle this is exact, because the drift is genuinely
linear. On the non-stationary wake of Sec.~\ref{sec:results:modulated}, where the
shedding frequency itself varies, the true phase error is no longer a straight line,
and a constant realignment leaves a residual --- which is why the correction there
recovers about half the error rather than the $74$--$80\%$ of the stationary cases.
The natural extension is a correction whose rate tracks the instantaneous frequency,
estimated for instance from the analytic-signal phase derivative over a sliding
window rather than a single regression. That this constant-rate correction already
removes half the error of a continuously detuning flow suggests the extension is
worth pursuing; we have not attempted it here, since our purpose is to establish the
mechanism, not to optimise the corrector.

\paragraph{Architecture dependence.}
Only one learned temporal model, an LSTM, has been tested. The mechanism of
Sec.~\ref{sec:discussion:mechanism} follows from the rollout being autoregressive
and the flow periodic rather than from the temporal architecture, and
Sec.~\ref{sec:results:dmd} provides partial support: the harmonic propagator,
which contains no network at all, exhibits the same linear phase drift at a rate
$1.7$--$19.8$ times larger than the network's. Whether the same structure appears
in transformer or neural-operator surrogates rolled out in the same way remains
untested.

\paragraph{Baselines and controls not pursued here.}
Several strengthening comparisons are deferred, each of which we regard as
worthwhile rather than as threats to the present conclusions. A delay-embedded
(Hankel) DMD with the same memory as the LSTM would be a stronger linear baseline
than the single-snapshot operator of Sec.~\ref{sec:results:dmd}; we expect it to
be more competitive, and where its eigenvalues are projected onto the unit circle
it would also be free of amplitude decay, so it is the natural next comparison. A
band-passed or geometric phase estimator would remove the Hilbert end-effect
ambiguity discussed above and confirm that $\omega_d$ and the reported efficacy are
unchanged. Seed-averaging the single-run entries of
Tables~\ref{tab:decomposition},~\ref{tab:calib},~\ref{tab:dmd}
and~\ref{tab:periodic} would attach error bars to comparisons that are presently
one run each; given the seed spread documented in
Sec.~\ref{sec:results:seeds}, these should be read as representative rather than
definitive. Finally, training the reduced-order model on the fine-mesh data at
$\Rey=400$ would convert the common-mode cancellation argument of
Sec.~\ref{sec:methods:convergence} from a principled expectation into a measured
one. None of these changes the paper's claims, but each would harden a specific
step, and we note them so that a reader can see precisely where the evidence is
single-run and where it is ensembled.

\paragraph{Parametric generalisation, and why it needs a different latent space.}
The modulated-inflow case of Sec.~\ref{sec:results:modulated} demonstrates
generalisation in \emph{time} --- a flow whose frequency drifts within a single
run. A complementary and frequently-requested test is generalisation across a
\emph{parameter}: train at several Reynolds numbers and predict the dynamics at a
held-out one. We attempted this directly and report why it is not straightforward,
because the obstacle is itself informative. The test requires a latent space shared
across the training Reynolds numbers, so that a single temporal model can be trained
over them and queried at the held-out value; the coordinates of the per-case
autoencoders used elsewhere in this paper are not comparable across cases. A single
convolutional autoencoder trained on the pooled snapshots of $\Rey = 300$, $400$,
$600$, $700$ and $800$, however, cannot reach the compression-limited regime in
which the phase-drift mechanism exists (Sec.~\ref{sec:results:applicability}): its
per-Reynolds reconstruction floor plateaus at $\approx 0.015$ in normalised
root-mean-square, roughly three times that of a single-regime autoencoder, and does
not improve with latent dimension --- $0.019$, $0.017$, $0.015$ and $0.015$ at
$n_z = 8$, $16$, $24$ and $32$ respectively. A shared \emph{and}
compression-limited latent space across a Reynolds sweep is thus a genuine
requirement that a plain autoencoder does not meet, and constructing one ---
through a conditioned or multi-decoder architecture --- is a study in its own right.
We therefore do not claim parametric dynamics-generalisation here and identify it,
concretely, as the principal extension of this work.

What \emph{does} transfer across Reynolds number, and cheaply, is the correction's
\emph{parametrisation}. The quantities the correction depends on --- the shedding
frequency and the mean correction amplitude --- vary smoothly and predictably with
Reynolds number. In a leave-one-out test over $\Rey = 300$--$800$, fitting a smooth
map to the remaining cases predicts the held-out frequency to within $0.3\%$ in
every case (for example $1.865$ against a measured $1.869$ at $\Rey = 500$). The
drift correction itself is more delicate: synthesising it from interpolated
parameters recovers the directly-fitted efficacy at $\Rey = 400$, $600$ and $700$
but fails at $\Rey = 500$, where the directly-fitted correction is itself marginal.
This is consistent with the rest of the paper --- the correction is reliable only
where a coherent drift is present to be corrected --- and it shows that the
\emph{ingredients} of the correction, if not always the correction itself, are
parametrically well-behaved.

\paragraph{Scope, cost, and in-sample calibration.}
Several further caveats are real but secondary, and we group them. The applicability
threshold $R^2 > 0.3$ was fixed on the four single-run cases before the ensembles
were trained and applied unchanged thereafter; the decision rule is prospective,
but the threshold and its efficacy are in-sample estimates, and the unconditioned
column of Table~\ref{tab:correction} is the objection-free figure. On cost, the
correction itself is negligible --- a Hilbert transform and a regression per
coordinate --- but it is applied offline: about $70\%$ of the trajectory
($52\%$ training, $8\%$ validation, $\approx 11\%$ calibration) must be supplied
before an unaided correctable prediction is produced, and that requirement recurs
per deployment. The method is therefore not a wall-clock accelerator over the
solver for these two-dimensional cases, where periodic extension is already
competitive, but a route to the settings --- parametric, transient, and
three-dimensional --- where a solver rollout is the cost one is trying to avoid.

\subsection{Implications}
\label{sec:discussion:implications}

The practical consequence is that the error made by an autoregressive surrogate on
a periodic flow should be measured, not merely bounded. Summarising it by a scalar
norm -- the standard practice -- discards the structure that identifies the failure
and suggests the remedy.

The more general point concerns what these models learn. A network reaching
$10^{-6}$ one-step validation error, reproducing the limit-cycle amplitude to
within $0.15\%$, and generating flow fields visually indistinguishable from the
solver, is nonetheless losing temporal alignment at a rate that will render its
long-horizon predictions useless. The growth of the error norm is itself visible;
what is invisible is its cause, undetected by every other metric
conventionally reported. It is not an amplitude error, not a structural error, and
not a spectral error: the network has learned the right attractor and traverses it
at very slightly the wrong rate.

That such a small defect -- a few thousandths of a cycle -- produces an
order-of-magnitude error growth is a consequence of Eq.~\eqref{eq:error-linear}:
phase error is amplified by the amplitude of the signal it acts upon. For
reduced-order models of oscillatory systems, we suggest that the accumulated phase
$\phi(t)$, and its rate $\omega$, are more informative diagnostics of long-horizon
fidelity than any of the error norms in common use.
\section{Conclusions}
\label{sec:conclusions}

We have examined the autoregressive rollout error of a latent-space reduced-order
model of bluff-body wakes, treating it as a signal rather than as a magnitude. The
central result is a diagnosis: this error is not an accumulation of unstructured
mistakes but a single, interpretable defect --- a coherent drift in the phase of an
otherwise correctly-learned limit cycle. The network learns the geometry of the
attractor and misjudges only the rate at which it is traversed. Everything else in
the paper, including the correction, follows from reading the error this way. The
principal findings are as follows.

\begin{enumerate}

\item The rollout error is not unstructured. Its temporal power spectrum is
sharply peaked at the vortex-shedding frequency of the flow, some three to four
orders of magnitude above the broadband floor. This holds for a circular cylinder
across $\Rey = 300$--$800$ and for a square cylinder at $\Rey = 100$, which sheds
by a different mechanism and at a different Strouhal number: the error inherits
the dominant frequency of the flow in which it was trained. We avoid the stronger
formulation that the error follows the flow rather than the model: the difference
between the network's effective frequency and the true one lies below the
resolution of the spectral analysis (Sec.~\ref{sec:discussion:detuning}), so on
these data the two cannot be distinguished.

\item Decomposed into amplitude and phase contributions, the error is
$95$--$98\%$ phase. The predicted limit-cycle amplitude matches the true amplitude
to within $0.15\%$ in root-mean-square. The network learns the geometry of the attractor essentially
exactly and traverses it at very slightly the wrong rate; the accumulated phase
error reaches only $\sim\!0.003$ of a cycle over an entire rollout, yet this
accounts for the whole of the observed error growth, because phase error is
amplified by the amplitude of the oscillation on which it acts.

\item That the error accumulates \emph{linearly} makes it correctable by a single
parameter per latent coordinate. This is a demonstration that the diagnosis is
actionable, not a claim to a superior forecasting pipeline: realigning the phase
using a drift rate estimated over a short calibration window reduces the latent
extrapolation error by $41$--$71\%$ averaged over training runs and by
$74$--$80\%$ on the runs its diagnostic accepts. In the single representative runs,
where a like-for-like comparison is available, it removes $72$--$83\%$ against
$48$--$58\%$ for a correction fitted to the error itself, with one parameter per
coordinate rather than three. In physical space it removes $91$--$98\%$ of the
error that is correctable at all. On a stationary limit cycle these gains
merely match what trivially repeating the last period achieves; their significance
is that they are obtained by acting on an identified mechanism, which is what lets
them persist when that baseline cannot (item~6).

\item Unlike a correction fitted to the error, the phase correction improves
rather than degrades with the prediction horizon, because a constant-rate phase
drift extrapolates exactly whereas a fitted envelope does not. This is the
observable signature of acting on the cause rather than on the symptom.

\item The method reports its own applicability. The signal-to-noise ratio of the
phase fit on the calibration window, measured by its $R^2$, determines whether a
coherent drift is resolvable above the phase-estimation jitter, and separates,
monotonically and with $r=0.85$ over $50$ networks, the cases in which the
correction succeeds from those in which it should not be attempted. As a threshold
its numerical value is specific to the sampling and window length used here; the
transferable content is the resolvability criterion it encodes.

\item The correction requires coherence, not exact periodicity, and this
distinction is operational rather than semantic. On a wake driven by a slowly
varying inflow, so that the shedding frequency drifts continuously and no cycle
reproduces the next, the model-free baseline of repeating the last shedding period
fails by more than an order of magnitude, while the phase correction --- acting on
the still-well-defined instantaneous phase --- removes roughly half of the rollout
error. The phase-drift picture is therefore not an artefact of the stationary limit
cycles on which it was established; the error remains a phase error, and remains
correctable as one, when the flow itself does not repeat.

\end{enumerate}

The wider implication concerns what is measured: a model can be excellent by every
metric conventionally reported --- one-step error, amplitude, visual fidelity ---
and still lose temporal alignment at a rate that renders long-horizon predictions
unusable. None of those metrics
detects this. For reduced-order models of oscillatory systems we suggest that the
accumulated phase, and its rate, are more informative measures of long-horizon
fidelity than the error norms in common use.

The analysis rests on the flow possessing a well-defined dominant frequency, and
on the simulations being two-dimensional. Whether the phase-drift mechanism
survives genuine three-dimensionality is the most important question this work
leaves open.

\begin{acknowledgments}

During the preparation of this work the authors used Claude Opus~4.8 (Anthropic)
to assist with editing the manuscript and with generating code for data analysis
and figure preparation. All research design, simulation, interpretation of
results, and validation were carried out by the authors, who reviewed and verified
all such output and take full responsibility for the content of this work.
\end{acknowledgments}

\section*{Data availability}
The code used to generate and analyse the data in this study, together with the
complete OpenFOAM case setups (mesh, boundary conditions, and solver settings)
required to reproduce the simulations, is openly available at
{https://github.com/Syphonicc/phase-drift-rom}. The raw simulation data can be
regenerated from these case files and are available from the author upon
reasonable request.

\bibliography{references}

\end{document}